\documentclass[a4paper,12pt]{article}
\usepackage{jheppub,esint,psfrag}
\usepackage[utf8]{inputenc}

\usepackage{epsfig,amssymb,amsmath,psfrag,subfigure,rotate,color,wasysym,xcolor}
\usepackage[bbgreekl]{mathbbol}
\usepackage[normalem]{ulem}
\usepackage{hyperref}

\allowdisplaybreaks 
\newcommand{\insertfig}[2]{\includegraphics[width=#1cm]{#2}}

\DeclareSymbolFontAlphabet{\mathbbm}{bbold}
\DeclareSymbolFontAlphabet{\mathbb}{AMSb}%

\def\XXint#1#2#3{{\setbox0=\hbox{$#1{#2#3}{\int}$ }
\vcenter{\hbox{$#2#3$ }}\kern-.6\wd0}}

\def \be  {\begin{equation}}
\def \ee  {\end{equation}}
\def \ba  {\begin{eqnarray}}
\def \ea  {\end{eqnarray}}
\def \baa {\begin{eqnarray*}}
\def \eaa {\end{eqnarray*}}
\newcommand{\ep}{\varepsilon}
\def \lab #1 {\label{#1}}

\newcommand\re[1]{(\ref{#1})}
\def\d{\hbox{{d}\kern-.20em\hbox{l}}}

\def \matrix #1 {\left(\begin{array}{cc} #1 \end{array}\right)}

\def \tr {\mathop{\rm tr}\nolimits}
\def \e  {\mathop{\rm e}\nolimits}

\newcommand \ket [1] {|{#1}\rangle}
\newcommand \bra [1] {\langle {#1}|}
\newcommand{\bit}[1]{\mbox{\boldmath$#1$}}
\def\1{\hbox{{1}\kern-.25em\hbox{l}}}

\newcommand{\ft}[2]{{\textstyle\frac{#1}{#2}}}

\makeatletter

\input pdf-trans
\newbox\qbox
\def\usecolor#1{\csname\string\color@#1\endcsname\space}
\newcommand\bordercolor[1]{\colsplit{1}{#1}}
\newcommand\fillcolor[1]{\colsplit{0}{#1}}
\newcommand\outline[1]{\leavevmode%
  \def\maltext{#1}%
  \setbox\qbox=\hbox{\maltext}%
  \boxgs{Q q 2 Tr \thickness\space w \fillcol\space \bordercol\space}{}%
  \copy\qbox%
}
\makeatother
\newcommand\colsplit[2]{\colorlet{tmpcolor}{#2}\edef\tmp{\usecolor{tmpcolor}}%
  \def\tmpB{}\expandafter\colsplithelp\tmp\relax%
  \ifnum0=#1\relax\edef\fillcol{\tmpB}\else\edef\bordercol{\tmpC}\fi}
\def\colsplithelp#1#2 #3\relax{%
  \edef\tmpB{\tmpB#1#2 }%
  \ifnum `#1>`9\relax\def\tmpC{#3}\else\colsplithelp#3\relax\fi
}
\bordercolor{black}
\fillcolor{white}
\def\thickness{.3}

\def\1{\mathbbm{1}}

\newcommand{\namedref}[2]{\hyperref[#2]{#1~\ref*{#2}}}
\newcommand{\secref}[1]{\namedref{Section}{#1}}
\newcommand{\appref}[1]{\namedref{Appendix}{#1}}

\newcommand{\figref}[1]{\namedref{Figure}{#1}}

\title{Walking Sudakov: From Cusp to Octagon}

\author[a]{Luis F. Alday,}
\author[b,c]{Elisabetta Armanini,}
\author[d]{Andrei V. Belitsky,}
\author[e]{Kelian H\"{a}ring,}
\author[b]{Alexander Zhiboedov}
\affiliation[a]{Mathematical Institute, University of Oxford, Andrew Wiles Building, Radcliffe Observatory Quarter, Woodstock Road, Oxford, OX2 6GG, U.K.}  
\affiliation[b]{CERN, Theoretical Physics Department, Geneva, Switzerland}
\affiliation[c]{Fields and Strings Laboratory, Institute of Physics,
\'Ecole Polytechnique F\'ed\'erale de Lausanne (EPFL), Route de la Sorge, CH-1015 Lausanne, Switzerland}
\affiliation[d]{Department of Physics, Arizona State University, Tempe, AZ 85287-1504, USA}
\affiliation[e]{Institute for Theoretical Physics, University of Amsterdam, 1090 GL Amsterdam, The Netherlands}

\abstract{We study the Sudakov form factor and the four-point scattering amplitude
on the Coulomb branch of planar ${\cal N}=4$ SYM as functions of the
Coulomb-branch parameters and kinematic invariants. This setup provides a
controlled probe of the interpolation between on- and off-shell regimes
of infrared-sensitive quantities in gauge theories. 
We identify a novel scaling limit in which both observables exhibit double-logarithmic behavior governed by a walking anomalous dimension.
As the mass scales are
varied, this walking anomalous dimension interpolates between the cusp anomalous dimension
of the on-shell regime and the octagon anomalous dimension of the
off-shell regime.
Based on the explicit two-loop result and the expected all-order
structure, we propose an all-loop form for the walking anomalous
dimension both for the form factor and for the four-point scattering amplitude. These all-loop expressions depend on new, presently unknown functions of the
't Hooft coupling. 
\vspace{3cm}}

\vspace{10cm}

\dedicated{
    \vspace{\fill}
    \begin{center}
        \itshape \sl Dedicated to George Sterman on the occasion of his Octogintennial
    \end{center}
    \vspace{\fill}
}

\begin{document}
\begin{flushleft}
\hfill \parbox[c]{40mm}{CERN-TH-2026-109}
\end{flushleft}
\maketitle

\newpage
\section{Introduction}

Precision predictions for high-energy processes in QCD rely on a robust understanding of double-logarithmic, or Sudakov, effects 
arising from soft and collinear momentum regions \cite{Mueller:1981sg,Sen:1982bt,Collins:1984kg,Collins:1989gx,Collins:1989bt,Sterman:1995fz,Agarwal:2021ais}. The simplest quantity in which this behavior is manifest is the
two-particle matrix element of the electromagnetic current, namely the
Sudakov form factor \cite{Sudakov:1954sw,Jackiw:1968zz,Fishbane:1971jz}. Besides being of intrinsic interest, the Sudakov form factor appears as a universal building block in factorization theorems 
\cite{Collins:1981uk,Sen:1982bt,Collins:1989gx,Sterman:1995fz,Feige:2014wja,Ma:2019hjq,Agarwal:2021ais,Ma:2026pjx} 
for quark and gluon scattering amplitudes \cite{Mueller:1981sg,Botts:1989kf,Mangano:1990by,Bern:1996je,Kidonakis:1997gm,Beneke:2009rj,Ferroglia:2009ii,Elvang:2013cua} 
by encoding their infrared behavior 
\cite{Sen:1982bt,Bassetto:1983mvz,Sterman:1986aj,Contopanagos:1996nh,Catani:1998bh,Sterman:2002qn,Bern:2005iz,Aybat:2006mz,Dixon:2008gr,Becher:2009qa,Gardi:2009qi,Dixon:2009ur,Almelid:2015jia,Gardi:2025lws}. 
Depending on specific circumstances, external states could be on or off their mass shell. How does the gauge dynamics differ in these two cases? Can one interpolate smoothly between them? These are the questions we address in this paper.

The on-shell massless Sudakov form factor has been the subject of intense studies since the early days of QCD 
\cite{Mueller:1979ih,Collins:1980ih,Sen:1982bt,Korchemsky:1986fj,Korchemsky:1988hd,Collins:1989bt,Magnea:1990zb,Magnea:2000ss}. 
Its double-logarithmic behavior is controlled by renormalization-group evolution
\cite{Ivanov:1985np,Korchemsky:1985xu,Korchemsky:1985xj,Korchemsky:1986fj,Korchemsky:1988hd}, whose leading coefficient is the cusp anomalous dimension 
\cite{Polyakov:1980ca,Dotsenko:1979wb,Korchemsky:1987wg,Kidonakis:2009ev,Grozin:2015kna,Kidonakis:2023lgc,Grozin:2022umo}.

Until very recently, much less was known about the off-shell regime of the Sudakov form factor, despite the fact that Sudakov's seminal original 
calculation was performed precisely in this regime \cite{Sudakov:1954sw}. There are several reasons for this. First, the off-shell form factor is less ubiquitous
in physical applications than its on-shell counterpart. Second, one must
address the practical problem of preserving gauge invariance for off-shell
matrix elements in order to define and compute them correctly. Third, loop
integrals in this regime receive dominant contributions from ultrasoft
momentum modes \cite{Mueller:1981sg}. In asymptotically free theories, this
can make the result sensitive to confinement-scale physics, thereby limiting
the applicability of perturbation theory from the outset.

In this paper, we study the interpolation between on- and off-shell regimes in the simplest four-dimensional gauge theory, namely planar maximally 
supersymmetric Yang--Mills theory, or $\mathcal{N} = 4$ SYM \cite{Brink:1976bc,Gliozzi:1976qd}. The theory is ultraviolet finite \cite{Mandelstam:1982cb}, 
and its coupling constant does not run. Nevertheless,
with possible future QCD applications in mind, we will assume that the ultrasoft momentum modes remain perturbative. 

The issue of gauge invariance is naturally addressed by working on the
Coulomb branch of the theory. This is achieved by giving vacuum expectation
values to some of the scalar fields \cite{Alday:2009zm}.
For external particles, the mass plays the role of a virtuality. More
precisely, in this work we analytically continue the standard Coulomb-branch
setup to complex angles on $S^5$ and work with Euclidean external virtualities. A further advantage of introducing masses for particles
propagating in loops is that the infrared divergences are regulated by these
masses, so that the full integrals can be defined in four dimensions.
The Sudakov form factor on the special Coulomb branch \cite{Caron-Huot:2021usw,Bork:2022vat}, where only external particles have 
masses, while all excitations propagating in quantum loops are strictly massless, was calculated to three-loop order in 
\cite{Belitsky:2022itf,Belitsky:2023ssv}. Similarly, the four-point scattering amplitude with only external nonzero masses was studied for example in \cite{Mandelstam:1959bc, Caron-Huot:2021usw}. The exponent controlling their
double-logarithmic asymptotics was shown to be the octagon anomalous
dimension \cite{Coronado:2018cxj,Belitsky:2019fan,Belitsky:2020qzm}, rather than the
cusp anomalous dimension. Thus the two limiting regimes are now well understood. In this work, we bridge them by allowing the internal and external masses to take different
values, thereby smoothly interpolating between the on- and off-shell
kinematic regimes. In this paper, we use the mostly-minus signature.

\subsection{Summary of the results}

The main result of the paper is the identification of a new high-energy limit in the two-particle Sudakov form factor and in the four-point scattering amplitude on the Coulomb branch of $\mathcal{N} = 4$ SYM. This high-energy limit generalizes the usual double-logarithmic behavior of the Sudakov form factor and of the scattering amplitude and is controlled by a new anomalous dimension that smoothly interpolates between the cusp and the octagon anomalous dimensions. For this reason we call it the walking anomalous dimension.

\begin{figure}[h!]
    \centering
    \includegraphics[width=0.8\linewidth]{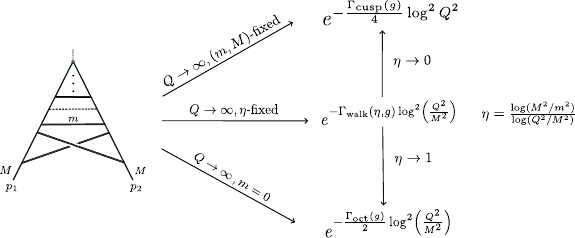}
    \caption{Different high-energy limits of the Sudakov form factor have different behaviors. The walking anomalous dimension in the middle interpolates between the well-known limits \eqref{OnShellSudakov} and \eqref{OffShellSudakov}. Here $M$ is the external Euclidean virtuality $p_1^2 = p_2^2 = -M^2$ and $Q$ is the Euclidean momentum transfer $Q^2 = -(p_1 + p_2)^2$. As the walking parameter $\eta \to 0$ the walking anomalous dimension reduces to the usual cusp-coefficient $\Gamma_{\text{cusp}}(g)/4$, while for $\eta \to 1$ it reduces to $\Gamma_{\text{oct}}(g)/2$.}
    \label{fig:walking_summary_ff}
\end{figure}

For the Sudakov form factor the different high-energy limits of \cite{Henn:2011by} and \cite{Belitsky:2022itf}, including our new high-energy limit are summarized in \figref{fig:walking_summary_ff}.\footnote{Equivalently, we can work in units where $Q=1$ and explore various `soft' limits $(m,M)\to 0$.}

\begin{figure}[h!]
    \centering
    \includegraphics[width=\linewidth]{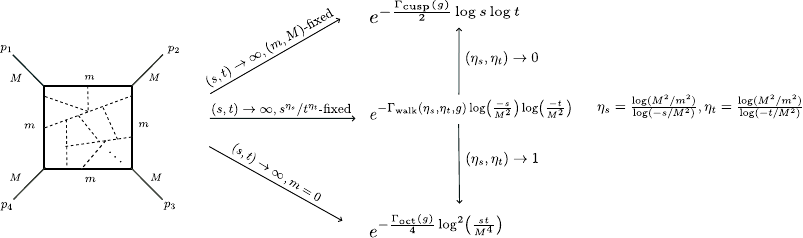}
    \caption{Different high-energy limits of the scattering amplitude have different behaviors. The walking anomalous dimension in the middle interpolates between the well-known limits \eqref{eq:fixed_angle} and \eqref{eq:fixed_angle_m0}. Here $M$ is the external Euclidean virtuality $p_i^2 = -M^2$ and $s = (p_1+p_2)^2, \, t = (p_2 +p_3)^2$.}
    \label{fig:walking_summary}
\end{figure}

The case of the four-point amplitude is richer since the amplitude depends on two Mandelstam variables $s$ and $t$, which can scale to infinity at different rates. 
The walking anomalous dimension depends then on two walking parameters $\eta_s,\eta_t$, as summarized in \figref{fig:walking_summary}. 

In the limit $\eta_s = \eta_t \equiv \eta$ the walking anomalous dimensions found in the form factor and in the amplitude agree up to a factor of $2$, generalizing the relation found in \cite{Belitsky:2025bgb},
\begin{equation}
\label{eq:satoff}
\Gamma_{\text{walk}}(\eta_s,\eta_t,g) \Big{|}_{\eta_s = \eta_t = \eta} = 2 \, \Gamma_{\text{walk}}(\eta,g) \ . 
\end{equation}

\section{The setup}
\begin{figure}[h!]
    \centering
    \includegraphics[width=0.8\linewidth]{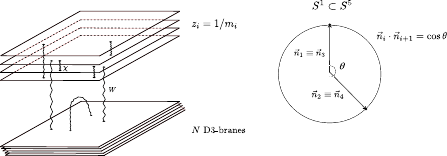}
    \caption{The relevant Coulomb branch setup for our analysis of the scattering amplitude with $k = 4$. For the form factor the setup is the same and it is enough to consider $k = 2$ separated D3-branes. The internal W-bosons have mass $m_i$ and the external mesons $\chi$ have mass $|m_i \vec{n}_i -m_{i+1} \vec{n}_{i+1}|$. Separating the probe D3-branes on $S^1 \subset S^5$ as depicted on the right, allows one to have massive external states even in the simple case $m_i \equiv m$ if $\theta \neq 0$. This is the case we will focus on.}
    \label{fig:setup_cb}
\end{figure}

We consider planar $\mathcal{N} = 4$ SYM on the Coulomb branch by turning on vacuum expectation
values for the six adjoint scalars
\[
\langle \Phi^I_{N+i,N+j} \rangle
=
\frac{\sqrt{2}}{g_{\mathrm{YM}}}\,
M^I_{N+i,N+j},
\qquad I=4,\ldots,9.
\]
This moves the theory away from the origin of moduli
space and Higgses the gauge symmetry to the subgroup
that commutes with the scalar background
\cite{Schabinger:2008,Alday:2009zm,Craig:2011}. In the sector relevant for us, we take the VEV matrix in
the $(N+i,N+j)$ block to be diagonal,
\begin{equation}
\vec{M}_{N+i,N+j}
= \text{diag}\left( m_1 \vec{n}_1,m_2 \vec{n}_2, \dots, m_k \vec{n}_k \right)\,,
\end{equation}
with $\vec{n}_i^{\,2}=1$, labeling a point on $S^5$. The parameters $m_i$ set the magnitudes of the
corresponding scalar VEVs, while the unit vectors
$\vec{n}_i$ determine their orientation in the internal
$SO(6)_R$ space.
The gauge group $U(N+k)$ is spontaneously broken to $U(N) \times U(1)^k$ and, from the perspective of the $AdS/CFT$ correspondence, this amounts to separating $k$ D3-branes from the stack and place them at fixed radial positions $z_i \sim 1/m_i$ of $AdS_5$. This is represented in \figref{fig:setup_cb} for the case $k = 4$, relevant for the four-point scattering amplitude.
For generic parameters, the off-diagonal fields $\chi_{i,j}$
associated with broken generators become massive BPS
states, with masses proportional to the distances
$|m_i\vec{n}_i - m_j\vec{n}_j|$ between the corresponding
points in $\mathbb{R}^6$
\cite{Craig:2011,Herderschee:2019}. These are the external states in our analysis and correspond to open strings stretched between the separated $k$ D3-branes. The internal states exchanged at leading order at large $N$ correspond to the massive W-bosons of mass $m_i$, and are strings stretched from the stack to the $k$ D3-branes.
We also separate the $k$ probe D3-branes along the directions of $S^5$ by placing them at different positions in $S^1 \subset S^5$. The R-symmetry configuration that we consider is
\begin{equation}
\vec{n}_i \cdot \vec{n}_{i+2} = 1\,, \quad \vec{n}_i \cdot \vec{n}_{i+1} = \cos(\theta)\,,
\end{equation}
which allows to have massive external states even in the simple case $m_i \equiv m$ for $\theta \neq 0$. This is precisely the situation considered in this paper.

\section{Walking in the Sudakov form factor}

In this section we study the Sudakov form factor on the Coulomb branch of $\mathcal{N} = 4$ SYM, with both internal and external massive particles. We analyze the high-energy behavior of this observable and identify an interpolating function between the cusp and the octagon anomalous dimensions, appearing as the coefficient of the double logarithm.
The exact treatment of the $(m,M)$-interplay, even at very large values of $Q \gg (m,M)$, is a difficult task. However, our main interest is in
the transient behavior of the leading coefficient accompanying the double logarithm and this allows significant simplifications of the required 
kinematic conditions. 

\subsection{The $(m,M)$ Sudakov form factor}
To define the $(m,M)$-dependent Sudakov form factor it is sufficient to separate $k = 2$ D3-branes in the setup of \figref{fig:setup_cb}.
The Sudakov form factor $\mathcal{F}_2$ for $\mathcal{N} = 4$ SYM with the $U(N+2)$ gauge group is defined by the transition matrix element
\begin{align}
(2 \pi)^4 \delta^{(4)} (q - p_1 - p_2) \mathcal{F}_2 = \int d^4 x \, \e^{i q \cdot x} \bra{0} \mathcal{O}_2 (x) \ket{p_1, p_2}
\, ,
\end{align}
of the half-BPS operator built from $Z$-scalar fields
\begin{align}
\label{BPSoperator}
\mathcal{O}_2 = \frac{1}{2 (N+2)} \tr_{\rm adj} Z^2
\, ,
\end{align}
and external meson states satisfying the on-shell condition
\begin{equation}
    p_1^2 = p_2^2 = (m_1 \vec{n}_1 - m_{2} \vec{n}_{2})^2 \,.
\end{equation}
We consider the simplest case in which all the Coulomb-branch parameters are equal $m_i \equiv m$ and the R-symmetry configuration is
\begin{equation}
\vec{n}_1 \cdot \vec{n}_2 = \cos\theta\,.
\end{equation}
We define the momentum transfer $Q$ as
\begin{equation}
Q^2 = -(p_1 + p_2)^2\,.
\end{equation}
Since $\vec{n}_1$ and $\vec{n}_2$ are not aligned when $\theta \neq 0$, the external states are not massless
\begin{equation}
p_1^2 = p_2^2 = 2m^2\bigl(1-\vec{n}_1 \cdot \vec{n}_{2}\bigr)
    = 4m^2 \sin^2\!\left(\frac{\theta}{2}\right)\,.
\end{equation}
All momentum invariants are analytically continued to the Euclidean values to avoid imaginary parts associated to particle production. We denote the positive Euclidean external virtuality $M^2 \equiv -p_i^2$ which, in terms of the $S^1$ angle, corresponds to the analytic continuation $\theta \to i \vartheta$. This analytic continuation has been previously studied for example in \cite{Correa:2012nk,Correa:2012hh,Gromov:2016rrp}.

\begin{figure}[h!]
\begin{center}
\mbox{
\begin{picture}(0,185)(230,0)
\put(0,0){\insertfig{16}{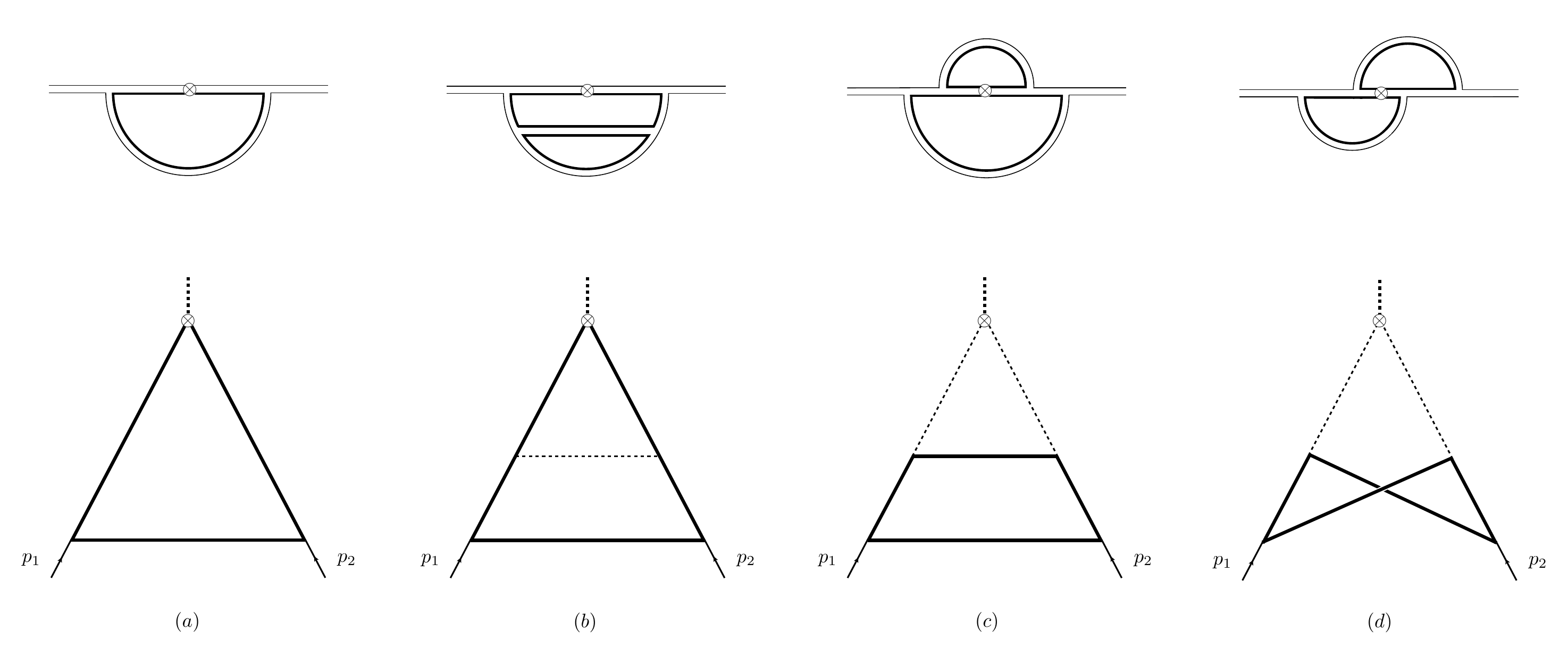}}
\end{picture}
}
\end{center}
\caption{\label{SudakovGraphs} The double-line representation and corresponding Feynman graphs for the one- and two-loop Sudakov form 
factor. In the double-line representation, the thick/thin lines correspond to the gauge groups $U(N) \times U(1)^2$ after spontaneous 
symmetry breaking, as explained in the main text. In Feynman graphs, the internal thick lines correspond to mass-$m$ propagators, while the 
dashed lines are massless. The external (thick solid) legs are ``virtual" with the off-shellness $M^2$.}
\end{figure}

Splitting up the $U(N+2)$ indices in the trace 
\re{BPSoperator} in terms of the unbroken $U(N)$ and broken $U(1)^2$ gauge groups, one encounters several graphs with various mass flows 
in quantum loops. This is demonstrated in Fig.\ \ref{SudakovGraphs} for the one- and two-loop contributions to the form factor 
in question.

Our goal in the next two sections is to calculate the Sudakov form factor to two-loop order 
\begin{align}
\mathcal{F}_2 = 1 + g^2 \mathcal{F}^{(1)}_2 + g^4 \mathcal{F}^{(2)}_2 + \dots
\end{align}
in 't Hooft coupling\footnote{We define it in $D = 4 - 2\varepsilon$ dimensions since our main computational strategy in the bulk of
this work will be based on the Method of Regions \cite{Beneke:1997zp}, which requires analytic continuation away from four 
space-time dimensions.}
\begin{align}
\label{DtHooft}
g^2 \equiv \e^{-\ep \gamma_{\rm\scriptscriptstyle E}} \frac{g_{\rm\scriptscriptstyle YM}^2 N}{(4 \pi)^{D/2}}
\, .
\end{align}
and study its functional form at asymptotically large values of the momentum transfer $Q$ and varying relative values of the
internal mass $m$ and external virtuality $M$. 

\subsection{From the cusp to the octagon}
\label{AnticipSection}
To start with, let us recall the known high-energy limits of the Sudakov form factor, highlighting its exponentiation property \cite{Henn:2011by}. The behavior of the form factor changes depending on which of the two masses is set to zero, with the remaining mass parameter small compared to $Q$.
For $M = 0$ and $Q \gg m$:
\begin{align}
\label{OnShellSudakov}
\log \mathcal{F}_2 = - \frac{\Gamma_{\rm cusp} (g)}{4} \log^2 \frac{Q^2}{m^2} + \Gamma_{\text{coll}}(g) \log \frac{Q^2}{m^2} + R_{\rm on}(g)\,.
\end{align}
The coefficient of the double logarithm is given by the cusp anomalous dimension, while the single-logarithm coefficient is the collinear anomalous dimension \cite{Polyakov:1980ca,Korchemsky:1987wg,Anastasiou:2003kj,Bern:2005iz}. The weak and strong coupling expansion of the cusp anomalous dimension reads
\begin{equation}
\Gamma_{\rm cusp} (g) = \begin{cases} 4 g^2 - 8 \zeta_2 g^4 + \mathcal{O}(g^6) \quad \text{for} \quad g \ll 1\,, \\
2 g -\frac{3 \log 2}{2 \pi} + \mathcal{O}(g^{-1}) \quad \text{for} \quad g \gg 1\,.
\end{cases}
\end{equation}
In the special case $m = 0$, a different high-energy behavior appears \cite{Belitsky:2022itf}
\begin{align}
\label{OffShellSudakov}
\log \mathcal{F}_2 = - \frac{\Gamma_{\rm oct} (g)}{2}  \log^2 \frac{Q^2}{M^2} + R_{\rm off} (g) + \mathcal{O}(Q^{-1})
\, ,
\end{align}
in terms of the octagon anomalous dimension \cite{Coronado:2018cxj,Belitsky:2019fan}
\begin{equation}
\Gamma_{\rm oct} (g) = \frac{2}{\pi^2} \log \cosh(2 \pi g) = \begin{cases} 4 g^2 - 16 \zeta_2 g^4 + \mathcal{O}(g^6) \quad \text{for} \quad g \ll 1\,, \\
\frac{4 g}{\pi} -\frac{2 \log 2}{\pi^2} + \mathcal{O}(g^{-1}) \quad \, \, \, \text{for} \quad g \gg 1\,.
\end{cases}
\end{equation}
The remainders $R_{\rm on}(g), R_{\rm off}(g)$ are kinematic-free functions of the 't Hooft coupling only \cite{Henn:2011by,Belitsky:2022itf}.\\

For general $(m,M)$, it is possible to consider a different high-energy limit, which will be explored in detail in the next sections. In particular, as we go to high-energies $Q \to \infty$, we can scale the ratio of the internal $m$ and external masses $M$. In terms of the dimensionless variables
\begin{equation}
\omega \equiv \frac{Q^2}{M^2}\,, \quad y \equiv \frac{m^2}{M^2}\,,
\label{eq:dimensionless_var_def}
\end{equation}
this new limit corresponds to $\omega \to \infty$, with 
\begin{equation}
\eta = -\frac{\log y}{\log \omega} \quad \text{fixed}\,.
\label{eq:definition_eta_v2}
\end{equation}
The Sudakov form factor in this limit exhibits a double-logarithmic behavior
\begin{equation}
\log \mathcal{F}_2 = -\Gamma_{\text{walk}}(\eta,g) \log^2 \omega + \mathcal{O}(\log \omega)\,,
\end{equation}
with $\Gamma_{\text{walk}}(\eta,g)$ the interpolating anomalous dimension computed in the following sections.

\subsection{Emergence of a new limit at one-loop}
\label{sec:one_loop_ff_1}

In this section, we analyze the one-loop Sudakov form factor starting from its dispersion representation. By a numerical and then analytical evaluation, we identify the interpolating anomalous dimension and the walking parameter $\eta$ which controls its behavior.\\

\begin{figure}[h!]
    \centering
    \includegraphics[width=0.85\linewidth]{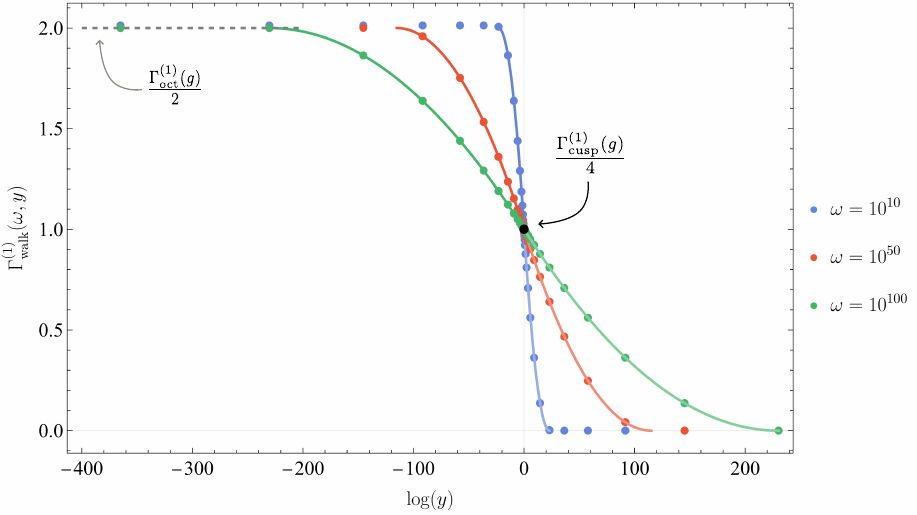}
    \caption{The one-loop walking anomalous dimension $\Gamma_{\text{walk}}^{(1)}(\omega,y) = \lim_{\omega \to \infty} (-\mathcal{F}_2^{(1)}/\log^2 \omega)$ obtained from the one-loop form factor \eqref{ExactOneLoopSudakov} as a function of $\log y \equiv \log (m^2/M^2)$ and different, large, values of $\omega$. The dots are obtained from the numerical evaluation of \eqref{ExactOneLoopSudakov}, while the curves are given by the explicit analytic expressions \eqref{Gwalk1} and \eqref{eq:Gwalk2}. We observe complete matching between the two for high enough energies.}
    \label{fig:walking_ff}
\end{figure}

At one loop, the form factor is given by the triangle graphs shown in Fig.\ \ref{SudakovGraphs}, and one can devise a simple
integral representation that treats its kinematics exactly. Namely, calculating the Cutkosky cut in the $Q$-channel, we find its
K\"all\'en-Lehmann dispersive representation in $D = 4$,
\begin{align}
\mathcal{F}_2^{(1)}
=
- Q^2
\int_{4 m^2}^\infty ds \frac{\rho (s)}{s + Q^2}
\, ,
\end{align}
with spectral density \cite{fronsdal1964integral,Norton:1964eeg}
\begin{align}
\rho (s) = \frac{2}{\sqrt{\lambda(s, -M^2, -M^2)}}
\log \left(
\frac{s + 2 M^2 + \sqrt{\lambda(s, -M^2, -M^2)} \sqrt{1 - 4 m^2/s}}{s + 2 M^2 - \sqrt{\lambda(s, -M^2, -M^2)} \sqrt{1 - 4 m^2/s}} \right)
\, , 
\end{align}
given in terms of the  K\"all\'en kinematic triangle function
\begin{align}
\lambda(x,y,z) = x^2 + y^2 + z^2 - 2 x y - 2 x z - 2 y z
\, .
\end{align}
In terms of the dimensionless variables defined before \eqref{eq:dimensionless_var_def} and with the change of variables $s = 4 M^2 (x+y)$, the one-loop form factor becomes
\begin{align}
\label{ExactOneLoopSudakov}
\mathcal{F}_2^{(1)}
=
- 4 \int_0^\infty dx 
\frac{
\log \frac{2 (x+y)+2 \sqrt{x} \sqrt{x+y+1}+1}{\sqrt{4 y (x+y+1)+1}}
}{
\sqrt{x+y} \sqrt{x+y+1} [1 + 4 (x+y)/\omega]
}
\, .
\end{align}
In order to show the emergence of the walking anomalous dimension, we first perform a numerical analysis of $\mathcal{F}_2^{(1)}$ in \eqref{ExactOneLoopSudakov}. Since we are interested in the coefficient of the double logarithm in the $\omega \to \infty$ limit, we numerically evaluate \eqref{ExactOneLoopSudakov} for large values of $\omega$ and then plot $(-\mathcal{F}_2^{(1)}/\log^2 \omega)$ as a function of $\log y$, which goes to $\Gamma_{\text{walk}}^{(1)}(\omega,y)$ as $\omega \to \infty$. The result of this numerical analysis is shown in \figref{fig:walking_ff}. It clearly demonstrates a walking behavior of $\Gamma^{(1)}_{\rm walk}(\omega, y)$ 
between the octagon and the cusp anomalous dimensions for $y < 1$. In particular, it reveals a shoulder forming for small values of $y \lesssim 1/\omega$, with the octagon limit 
being reached earlier for lower values of $\omega$. For $y > 1$ the anomalous dimension interpolates between the cusp and zero. 

\begin{figure}[h!]
    \centering
    \includegraphics[width=0.85\linewidth]{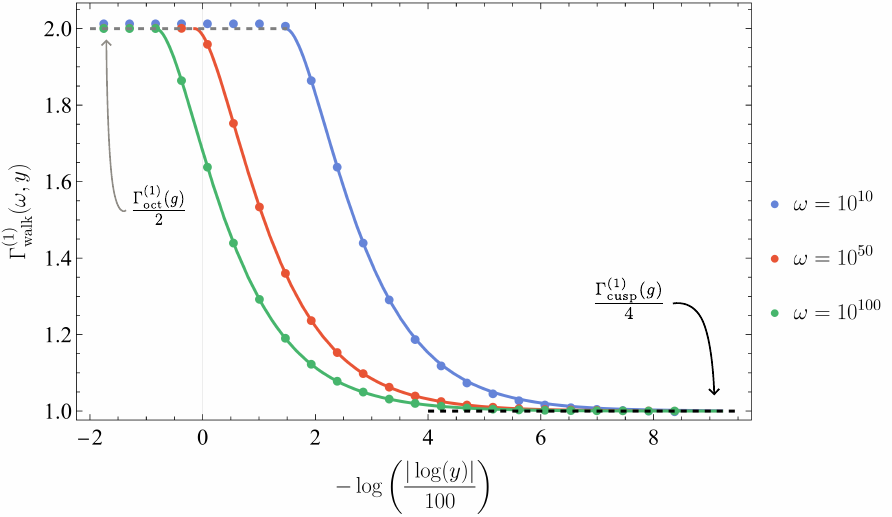}
    \caption{Zoom of the octagon-to-cusp-transition region ($y < 1$) for one-loop walking anomalous dimension $\Gamma_{\text{walk}}^{(1)}(\omega,y) = \lim_{\omega \to \infty} (-\mathcal{F}_2^{(1)}/\log^2 \omega)$. The plot is obtained by numerically evaluating \eqref{ExactOneLoopSudakov} as a function of $-\log\left(\frac{|\log y|}{100}\right) \equiv -\log\left(\frac{|\log (m^2/M^2)|}{100}\right)$ at different, large, values of $\omega$. The curve is the analytic expression \eqref{Gwalk1}.}
    \label{fig:walking_ff_zoom}
\end{figure}

Let us understand the octagon-to-cusp transition happening for $y < 1$ analytically. The plot in \figref{fig:walking_ff_zoom} focuses on the transition region $y < 1$ and highlights the octagon and the cusp endpoints which are smoothly connected by the walking anomalous dimension.

In particular, the transition occurs over the interval of $y \in [1/\omega, 1]$, so it suffices to formally consider the limit $y \ll 1$ and ignore the $y$-dependence
everywhere in the integrand of \eqref{ExactOneLoopSudakov}, except for the square root in the denominator of the logarithm. By rationalizing the integration variable with 
the substitution $x = t^2/(1 + 2 t)$, we find the following representation
\begin{align}
\mathcal{F}_2^{(1)}\Big{|}_{y \ll 1}
=
-
\int_0^\infty dt
\frac{\log \frac{(1 + t)^6}{[1 + t + (2 + t)^2 y]^2}}{(1+t) + t^2/\omega}
\, ,
\end{align}
which can be easily evaluated in terms of dilogarithms. This expression clarifies the shoulder structure of the walking Sudakov form 
factor. Namely, since the shoulder forms at $y = 1/\omega$, we need to analyze the integral separately for $y< 1/\omega$ and $y>1/\omega$.
For the behavior on the shoulder $y< 1/\omega$, we find
\begin{align}
\Gamma^{(1)}_{\rm walk}(y< \omega^{-1})
=
2 + \frac{4 \zeta_2}{\log^2 \omega} + y 
\left(
\frac{2 \omega \log y}{\log^2 \omega} - \frac{2 \omega}{\log^2 \omega}
-
\frac{4}{\log^2 \omega}+\frac{2 \omega}{\log \omega} - \frac{4}{\log \omega}
\right)
+
\mathcal{O}(y^2)
\, ,
\end{align}
while for the walking region $y > 1/\omega$, we get
\begin{align}
\label{LowerGamma}
\Gamma^{(1)}_{\rm walk}(y> \omega^{-1})
=
1 - \frac{2 \log y}{\log \omega} - \frac{\log^2 y}{\log ^2 \omega} 
+
y \frac{4 \log y}{\log ^2 \omega} 
+
\mathcal{O}(y^2)
\, .
\end{align}
Ignoring $\mathcal{O}(y)$ and higher-order terms in 
\re{LowerGamma}, we observe the emergence of the walking parameter $\eta$ as defined in \eqref{eq:definition_eta_v2},
which varies in the interval $\eta \in [0,1]$ for $1/\omega \leq y \leq 1$. In its terms, 
the walking one-loop anomalous dimension becomes
\begin{align}
\label{Gwalk1}
\Gamma^{(1)}_{\rm walk}(\eta) = 1 + 2 \eta - \eta^2
\,, \quad \eta \in [0,1]\,.
\end{align}
It has the one-loop octagon and the one-loop cusp anomalous dimensions as its boundary values
\begin{equation}
\Gamma^{(1)}_{\rm walk}(0) 
= \frac{\Gamma^{(1)}_{\rm cusp}}{4} \, , \qquad \Gamma^{(1)}_{\rm walk}(1)  =
\frac{\Gamma^{(1)}_{\rm oct} }{2}\, .
\end{equation}
For $\eta \geq 1$, the walking anomalous dimension is simply given by the octagon anomalous dimension
\begin{equation}
\Gamma_{\rm walk}^{(1)}(\eta \geq 1) = \frac{\Gamma_{\rm oct}^{(1)}}{2}\,.
\end{equation}
This can be expected by the fact that, for $\eta \geq 1$, $m^2 \leq M^4/Q^2$, so $m$ is below the relevant scale $M^2/Q$ for the off-shell Sudakov case \eqref{OffShellSudakov} and cannot be resolved at double-logarithmic accuracy.\\

The analytic expression of the walking anomalous dimension for $ y > 1$ can be found by considering the opposite $y \gg 1$ limit in the integrand of \eqref{ExactOneLoopSudakov}. By analyzing separately the case $ y < \omega$ and $y > \omega$, we find
\begin{equation}
\Gamma_{\text{walk}}^{(1)}(1 < y < \omega) = \left(1-\frac{\log y}{\log \omega}\right)^2 + \mathcal{O}(y^{-1})\,,
\label{eq:interpolation_neg_eta}
\end{equation}
while for the shoulder $y > \omega$, $\Gamma_{\text{walk}}^{(1)}(y > \omega) = 0$. This reflects the fact that, in the limit $m^2 \gg Q^2$, the loop degrees of freedom are not dynamical. In terms of the parameter $\eta$, the walking anomalous dimension for $\eta \in [-1,0]$ for $1 \leq y \leq \omega$ it is simply
\begin{equation}
\Gamma^{(1)}_{\rm walk}(\eta) = (1+\eta)^2
\,, \quad \eta \in [-1,0]\,.
\label{eq:Gwalk2}
\end{equation}
At one-loop the Sudakov walking anomalous dimension is
\begin{equation}
\Gamma_{\rm walk}^{(1)}(\eta) = g^2
\begin{cases}
0 \quad \text{for} \quad \eta \leq -1\,,\\
(1 + \eta)^2 \quad \text{for} \quad -1 \leq \eta \leq 0\,, \\
(1+2 \eta-\eta^2) \quad \text{for} \quad 0 \leq \eta \leq 1\,, \\
2 \quad \text{for} \quad \eta \geq 1\,,
\end{cases}
\label{eq:ff_one_loop_all}
\end{equation}
and is a $C^1$ function of the walking parameter $\eta$. 
In order to generalize these findings to higher loops, we will use a more systematic method, as explained in the next section.

\subsection{Two-loop computation with the Method of Regions}
\label{MofRCalcSection}

To calculate the walking anomalous dimension at two-loops, we will rely on the {\sl tropical} Method of Regions \cite{Belitsky:2025sin}. 
It is particularly well-suited for the construction of a systematic expansion in the logarithms of the small parameter $1/\omega$ with their 
accompanying coefficients determined by a sum of {\sl locally} finite integrals via Salvatori's  `subtraction' scheme \cite{Salvatori:2024nva}. 
The resulting integrals can then be straightforwardly handled with Panzer's {\tt HyperInt} integrator \cite{Panzer:2014gra}.

We begin with setting up our conventions. The first two perturbative orders of the Sudakov form factors are determined by the following 
four momentum integrals \cite{Henn:2011by}
\begin{align}
\mathcal{F}_2^{(1)} = - 2 Q^2 T_{11}
\, , \qquad
\mathcal{F}_2^{(2)} = Q^4 \Big( 2 T_{21,b} + 2 T_{21,c} + T_{22} \Big)
\, ,
\end{align}
shown in Fig.\ \ref{SudakovGraphs}, graphs $(a)$, $(b)$, $(c)$, and $(d)$, respectively. They read, explicitly, 
\begin{align}
\label{T11}
T_{11} [1,1,1]
&
= 
\int_k D_m (k) D_m (k+p_1) D_m (k-p_2)
\, , \\
\label{T21b}
T_{21,b} [1,1,1,1,1,1]
&
= 
\int_{k_1} \int_{k_2} D_m (k_1) D_0 (k_2) D_m (k_1 + p_1) 
\\
&
\times
D_m (k_1 + k_2 + p_1) D_m (k_1 - p_2) D_m (k_1 + k_2 - p_2) 
\, , \nonumber\\
\label{T21c}
T_{21,c} [1,1,1,1,1,1]
&
= 
\int_{k_1} \int_{k_2} D_m (k_1) D_m (k_2) D_m (k_1 + p_1) 
\\
&
\times
D_{\textcolor{black}{0}} (k_1 + k_2 + p_1) D_{\textcolor{black}{m}} (k_1 - p_2) D_0 (k_1 + k_2 - p_2) 
\, , \nonumber\\
\label{T22}
T_{22} [1,1,1,1,1,1]
&
= 
\int_{k_1} \int_{k_2} D_m (k_1) D_m (k_2) D_m (k_1 + p_1) 
\\
&
\times
D_0 (k_1 + k_2 + p_1) D_0 (k_1 + k_2 - p_2) D_m (k_2 - p_2) 
\, , \nonumber
\end{align}
with integrands given by products of the mass-$m$ propagators $D_m^{-1} (k) \equiv - k^2 + m^2$. The labels in the square brackets on 
the left-hand side of these equations designate the powers of Feynman propagators. The Method of Regions \cite{Beneke:1997zp}, 
to be employed in their calculation, partitions the entire momentum space in terms of individual momentum modes that possess certain 
scaling behavior with respect to the small parameter $1/\omega$. However, since this procedure is akin to the operator product expansion 
at a diagrammatic level, it inevitably introduces an approximation of corresponding integrands. It therefore generates singularities 
when these are integrated over an unconstrained momentum space. This calls for the introduction of an intermediate regularization, 
typically a dimensional regularization. That is why the above momentum integrals are defined with the loop integration measure taken 
in $D = 4 - 2 \varepsilon$ dimensions\footnote{Everywhere below, we set the scale parameter $\mu$ of dimreg to 1, i.e., $\mu = 1$.},
\begin{align}
\int_k \equiv 
{\rm e}^{\varepsilon \gamma_{\rm E}}
\int \frac{d^D k}{i \pi^{D/2}}
\, .
\end{align}
Notice that dimensional regularization alone will not suffice to make individual leading regions well-defined. We will be forced
to introduce an additional analytic regularization \cite{Speer:1968qxh} of the powers of Feynman propagators. This is the reason that 
we have to make these explicit in their definitions. 

Last but not least, since the form factor is a function of two dimensionless ratios \eqref{eq:dimensionless_var_def}, without loss of generality, we will set $Q^2 = 1$.

\subsubsection{A detailed one-loop analysis}
\label{1Lsection}

Let us proceed with a detailed analysis of the one-loop case, which can be performed by elementary means. We consider the asymptotic 
behavior of $T_{11}$ in \eqref{T11} in the limit 
\begin{align}
\label{KinematicsEq}
{\rm lim}
\equiv \{
\omega^{-1} \to 0
\, ,
y = \mbox{fixed} \neq 0
\}
\, ,
\end{align}
and immediately reveal \cite{Smirnov:2021rhf} three leading contributions with region vectors $\bit{r}_{\rm h} = (0,0,0)$, $\bit{r}_{\rm c1} = (0,1,0)$, 
and $\bit{r}_{\rm c2} = (0,0,1)$, corresponding to the hard and collinear scalings of the loop momentum, respectively. The latter two correspond
to the loop momentum being collinear to the momentum of the leg with momentum either $p_1$ or $p_2$. Dimensional regularization alone is unable to make these leading momentum-region contributions finite \cite{Collins:1992tv}. This is an artifact of rapidity divergences that emerge from the kinematic approximation employed to extract leading power effects \cite{Collins:2011zzd}. We therefore introduce a rapidity cut-off, equivalently implemented by an introduction of an analytic regulator \cite{Smirnov:1997gx,Becher:2011dz} for the two propagators connected with the hard momentum vertex, such that
\begin{align}
T_{11} \equiv T_{11} [1, 1 + \lambda, 1 - \lambda]
\, .
\end{align}
The sum of all regularized indices must coincide with the original sum, so that one does not imbalance the rescaling weight of the integrand.
Thus, at one loop
\begin{align}
T_{11}|_{\rm lim} 
= 
T_{11}^{\rm h} + T_{11}^{\rm c1} + T_{11}^{\rm c2}
\, ,
\end{align}
where, 
\begin{align}
T_{11}^{\rm h} 
&
=
- \frac{\e^{\gamma_{\rm E}  \varepsilon} \Gamma (-\varepsilon)^2 \Gamma (\varepsilon+1)}{\Gamma (1-2 \varepsilon)}
\, , \\
T_{11}^{\rm c1} 
&
=
\frac{\e^{\gamma_{\rm E}  \varepsilon} \Gamma (\varepsilon-\lambda) }{\Gamma (1-\lambda)}
\omega^{\varepsilon - \lambda}
\int_0^1 dt\,
t^{-1 -2 \lambda}
\Big(y + (1-t)t\Big)^{\lambda-\varepsilon}
\, , \\
T_{11}^{\rm c2} 
&
=
\frac{\e^{\gamma_{\rm E} \varepsilon} \Gamma (\varepsilon+\lambda) }{\Gamma (1+\lambda)}
\omega^{\varepsilon + \lambda}
\int_0^1 dt\,
t^{-1 +2 \lambda}
\Big(y + (1-t)t\Big)^{-\lambda-\varepsilon}
\, .
\end{align}
In their sum, the analytic regularization is removed first, $\lambda \to 0$, followed by $\varepsilon \to 0$. Notice that the Method of Regions is `blind' to zero-bin effects \cite{Manohar:2006nz,Dixon:2008gr} that reflect overlaps of various momentum regions when regularizations other than dimensional are employed in perturbative analyses. In our consideration, the effect of soft modes with momentum scalings $k_{\rm s} \sim m$ yield scaleless momentum integrals and are set identically to zero. So the collinear regions effectively include soft-collinear modes.

We extract the $1/\lambda$-poles from the collinear regions and cancel them first. This is accomplished by an appropriate analytic continuation 
in $\lambda$ and the use of a standard subtraction resulting in the $+$ prescription\footnote{The plus-prescription is defined as 
$$\int_0^1 dt \, f(t)/[t]_+ \equiv \int_0^1 dt \, \big(f(t) - f(0)\big)/t$$.}
\begin{align}
\int_0^1 \frac{dt}{t^{1\pm 2 \lambda}} f(t)
=
\mp \frac{f(0)}{2 \lambda}
+
\int_0^1 \frac{dt}{[t]_+} f(t)
+
O (\lambda)
\, . 
\end{align}
We then find
\begin{align}
\label{CollinearSfixedY}
T_{11}^{\rm c1} 
=
-
\e^{\gamma_{\rm E} \varepsilon} (y/\omega)^{\lambda-\varepsilon} 
\frac{\Gamma (\varepsilon-\lambda)}{2 \lambda \Gamma (1-\lambda)}
+
\e^{\gamma_{\rm E}  \varepsilon}  \Gamma (\varepsilon) \omega^{\varepsilon}
\int_0^1
\frac{dt}{t}
\Big(
(y + (1-t) t)^{-\varepsilon}
-
y^{-\varepsilon}\Big)
\, ,
\end{align}
and the same for $T_{11}^{\rm c2} $ with the obvious substitution $\lambda \to -\lambda$.

Subsequent manipulations depend on the {\sl order of limits}. For kinematics in Eq.\ (\ref{KinematicsEq}), the $t$-integral is {\sl locally finite} 
and we can safely Taylor expand in $\varepsilon$ the integrand itself. We find
\begin{align}
\label{YdepCollReg}
\int_0^1
\frac{dt}{t}
\Big(
(y + (1-t) t)^{-\varepsilon}
-
y^{-\varepsilon}\Big)
=
\varepsilon 
\left(
\text{Li}_2
\bigg(\frac{2}{1 - \sqrt{4 y+1}}\bigg)
+
\text{Li}_2\bigg(\frac{2}{1 + \sqrt{4 y+1}}\bigg)
\right)
+
O (\varepsilon^2)
\, .
\end{align}
This expression is safe to use for very large values of $y$ as well. It decays inversely as $1/y$. Depending on the value of $y$,
there is an order of limits issue for small $y$: $\varepsilon \to 0$ does not commute with $y \to 0$. Namely, the above 
calculation yields
\begin{align}
\label{SmallY}
\lim_{y \to 0}\lim_{\varepsilon \to 0}
\int_0^1
\frac{dt}{t}
\Big(
(y + (1-t) t)^{-\varepsilon}
-
y^{-\varepsilon}\Big)
=
-\frac{1}{2} \varepsilon \log^2 y
\, ,
\end{align}
as $y \to 0$. In the opposite order of limits, assuming analytic continuation to $\Re{\rm e} [\varepsilon] < 0$, we find
\begin{align}
\label{SmallYorder}
\lim_{\varepsilon \to 0} 
\lim_{y \to 0}
\int_0^1
\frac{dt}{t}
\Big(
(y + (1-t) t)^{-\varepsilon}
-
y^{-\varepsilon}\Big)
=
\lim_{\varepsilon \to 0}
\frac{\Gamma (1-\varepsilon) \Gamma (-\varepsilon)}{\Gamma (1-2 \varepsilon)}
\, ,
\end{align}
which develops a pole in $\varepsilon$. At the same time, we can neglect the first term in (\ref{CollinearSfixedY}) since it vanishes at 
small $y$ after analytic continuation in $\varepsilon$ to $\Re{\rm e} [\pm \lambda - \varepsilon] > 0$.

Let us add up these separate regions together for various values of $y$: large $1 \ll y \lesssim \omega$, walking $1/\omega \lesssim y \ll 1$, and
ultrasmall $y \ll 1/\omega$.
\begin{itemize}
\item {\bf Large $1 \ll y \lesssim \omega$}: Neglecting \eqref{YdepCollReg}, for the sum of the collinear regions we find
\begin{align}
\Big(T_{11}^{\rm c1} + T_{11}^{\rm c2} \Big)_{y \gg 1}
=
-
\frac{1}{\varepsilon^2}
+
\frac{1}{2} 
\left(
\log^2 (\omega/y) + \zeta_2
\right)
\, .
\end{align}
Combining it with the hard region,
\begin{align}
T_{11}^{\rm h}
=
\frac{1}{\varepsilon^2}
-
\frac{\zeta_2}{2} 
\, ,
\end{align}
we get 
\begin{equation}
\Big(T_{11}^{\rm c1} + T_{11}^{\rm c2} + T_{11}^h \Big)_{y \gg 1} = \frac{1}{2} \log^2(\omega/y)\,,
\label{eq:walking_neg_eta}
\end{equation}
which precisely gives the behavior for $-1 \leq \eta \leq 0$ found in \eqref{eq:interpolation_neg_eta}, recovering the on-shell one-loop Sudakov form factor \eqref{OnShellSudakov}.

\item {\bf Walking $y < 1$}: This is the walking region $\eta \in [0,1]$ of \secref{sec:one_loop_ff_1}, that we would like to reproduce. Thus, we 
have to add \re{SmallY} to the above large-$y$ expression, which is, in fact, valid for any value of $y$ as can be seen from the structure of the first term in \re{CollinearSfixedY}. 
\begin{align}
\Big(T_{11}^{\rm h}+T_{11}^{\rm c1}+T_{11}^{\rm c2}\Big)_{1/\omega \lesssim y \ll 1}
=
\frac{1}{2} 
\left(
\log^2 (\omega/y)- 2 \log^2 y
\right)
\, .
\label{eq:walking_pos_eta}
\end{align}
This is exactly the result of \eqref{Gwalk1}. This short consideration confirms the simplifying assumption we sought:
to uncover the walking behavior of the Sudakov form factor, it suffices to consider $y \ll 1$ only to extract the leading double logarithms. The deviation from the moderate values of $y$ manifests itself only in subleading logarithms, as we will explicitly see 
in the next section.

\item {\bf Ultrasmall $y$}: Using the properly interchanged order of limits in \eqref{SmallYorder}, we find for the sum of the 
hard and collinear regions
\begin{align}
\Big(T_{11}^{\rm h}+T_{11}^{\rm c1}+T_{11}^{\rm c2}\Big)_{y \ll 1/\omega}
=
-\frac{1}{\varepsilon^2}
-
\frac{1}{\varepsilon} \log \omega
- 
\log^2 \omega 
+
\frac{\zeta_2}{2}
\, .
\end{align}
We observe that the result possesses poles in $\varepsilon$, even though the original integral was finite. The reason for this is 
well-known, see, e.g., Ref.\ \cite{Becher:2014oda}. We made an assumption in the initial integral that $y \neq 0$ and used the 
Method of Regions to detect leading contributions. Thus, taking the limit $y \to 0$ falls outside its domain of applicability. To get 
out of this predicament, we have to relax the above constraint and allow for $y = 0$. This gives an extra leading, the so-called 
{\sl ultrasoft}, region with the loop momentum scaling as $k_{\rm us} \sim M^2/Q$ \cite{Fishbane:1971jz,Mueller:1979ih,Korchemsky:1988hd}. 
It reads
\begin{align}
\Big(T_{11}^{\rm us}\Big)_{y \ll 1/\omega}
=
\e^{\gamma_E  \varepsilon} \omega^{2 \varepsilon} \Gamma (1-\varepsilon) \Gamma (\varepsilon)^2
\, .
\end{align}
Adding it to the other three
\begin{align}
\Big(T_{11}^{\rm h}+T_{11}^{\rm c1}+T_{11}^{\rm c2}+T_{11}^{\rm us}\Big)_{y \ll 1/\omega}
=
\log^2 \omega + 2 \zeta_2
\, ,
\end{align}
we reproduce the one-loop off-shell result (\ref{OffShellSudakov}).
\end{itemize}

\subsubsection{Two-loop analysis and exponentiation}

Having established the kinematic domain reflecting the walking behavior of the Sudakov form factor, we now proceed to two loops. This will be done in two steps. First, we utilize {\tt FIESTA} to identify all nontrivial momentum regions as $1/\omega \to 0$. Again, to render individual contributions well-defined, an analytic regulator needs to be employed to tame rapidity divergences. However, if, as in the one-loop case, it is taken as an independent parameter $\lambda$, subsequent application of the tropical `subtraction' scheme \cite{Salvatori:2024nva}  will not be possible in its original, unaltered form. To alleviate this problem, we will make $\lambda$ proportional to $\varepsilon$ with relative coefficients being just integers. Thus, we define the two-loop momentum integrals as follows\footnote{Any choice of the analytic regulators is valid as long as one does not change rescaling properties of parametric integrals. Had we used $\lambda$ instead of $\varepsilon$, these would cancel in Laurent expansion even before the $\varepsilon$-expansion is employed.}
\begin{align*}
T_{21} 
&\equiv 
T_{21} [1, 1, 1 + 2 \varepsilon, 1 + 3 \varepsilon, 1 - 2 \varepsilon, 1 - 3 \varepsilon]
\, , \\
T_{22} 
&\equiv 
T_{22} [1 + 5 \varepsilon, 1 - 5 \varepsilon, 1 + 3 \varepsilon, 1 - 2 \varepsilon, 1 - 2 \varepsilon, 1 - 3 \varepsilon]
\, .
\end{align*}

Second, parametric Feynman integrals stemming from each non-vanishing region are then recast as a sum of {\sl locally finite} integrals \cite{Belitsky:2025sin,Salvatori:2024nva} such that one can perform their $\varepsilon$-expansion directly at the level of the {\sl integrand}. This procedure was discussed at length in \cite{Salvatori:2024nva,Belitsky:2025sin}, and it will not be repeated 
here. Before the final integration step, the exact $y$-dependence is rather complicated. While in many cases, results can be obtained 
in terms of Goncharov polylogarithms \cite{Goncharov:2001iea,Goncharov:2009lql} of weight up to four, with their arguments being 
algebraic functions of $y$, we did not succeed in this endeavor for all integrals. We observed a violation of linear reducibility of nested 
integrals early on in the integration process: it potentially points toward an elliptic nature of emerging functions. Since our goal is to merely 
uncover the walking behavior only for the leading double logarithms, we imposed the simplifying assumption that $y \ll 1$ or $y \gg 1$. 
{\tt HyperInt} \cite{Panzer:2014gra} integrates everything in terms of powers of $\log y$. The above strategy produces the following results
\begin{itemize}
\item Case $y \ll 1$, which allows us to compute\footnote{We are deeply indebted to Francesco Calisto, Subramanya Hegde, Martín Lagares, 
and Elia Mazzucchelli for sharing with us their upcoming paper \cite{Calisto2026} that hinted a potential error in the subleading logarithms. 
Indeed, we reanalyzed the non-planar ladder and corrected coefficients of subleading logarithms accordingly.} $\Gamma_{\rm walk}^{(2)}(\eta)$ for 
$\eta \in [0,1]$:
\begin{align}
T_{21, b}
&
=
\frac{1}{24} \log^4 \omega 
-
\frac{1}{6} \log^3 \omega \log y
+
\log^2 \omega \left( \frac{1}{4} \log^2 y + 2 \zeta_2 \right) +
\\
&
+
\frac{1}{6} \log \omega \log^3 y
-
\frac{1}{24} \log^4 y 
-
\zeta_2 \log^2 y
-
2 \zeta_3 \log y
+
4 \zeta_4
\, , \nonumber\\
T_{21, c}
&=
\frac{1}{24} \log^4 \omega
-
\frac{1}{6} \log^3 \omega \log y
+
\log^2 \omega 
\left(
\frac{1}{4} \log^2 y
+
2 \zeta_2
\right) +
\\
&
+
\log \omega 
\left(
\frac{1}{6} \log^3 y
-
2 \zeta_3 
\right)
-
\frac{1}{24} \log^4 y
-
\zeta_2 \log^2 y 
-
4 \zeta_3 \log y
+
\frac{9}{2} \zeta_4
\, , \nonumber\\
\label{T22ySmall}
T_{22}
&=
\frac{1}{3} \log^4 \omega 
-
\frac{4}{3} \log^3 \omega \log y
-
6 \zeta_2 \log^2 \omega +
\\
&
+
\log \omega 
\left(
\frac{4}{3} \log^3 y -  \textcolor{black}{4} \zeta_2 \log y +  \textcolor{black}{8} \zeta_3
\right)
+
\frac{2}{3} \log^4 y
+
 \textcolor{black}{6} \zeta_2 \log^2 y
+
 \textcolor{black}{8} \zeta_3 \log y
\mathbin{ \textcolor{black}{-}}
 \textcolor{black}{9} \zeta_4
\, . \nonumber
\end{align}
The sum of the one-loop \eqref{eq:walking_pos_eta} and two-loop results exponentiates and the logarithm of the Sudakov form factor exhibits at most double-logarithmic behavior
\begin{equation}
\begin{aligned}
\log \mathcal{F}_2
=
& - g^2 
\left( \log^2 \omega - 2 \log \omega \log y - \log^2 y  \right) \\
& +
g^4 
\left(
\zeta_2 \left( 2 \log^2 \omega -  \textcolor{black}{4} \log \omega \log y +  \textcolor{black}{2} \log^2 y \right)
+
 \textcolor{black}{4} \zeta_3 ( \log \omega \mathbin{ \textcolor{black}{-}} \log y )
+
 \textcolor{black}{8}
\zeta_4
\right)\,.
\label{eq:small_y_2loop}
\end{aligned}
\end{equation}
The two-loop correction to the walking anomalous dimension can be extracted from the leading double-logarithmic contribution
\begin{equation}
\Gamma^{(2)}_{\rm walk}(\eta)
= 
-  \textcolor{black}{2 \zeta_2 (1 + \eta)^2}
\,, \quad \eta \in [0,1]\,.
\end{equation}

\item Case $1 \ll y \lesssim \omega$, which allows to compute $\Gamma_{\rm walk}^{(2)}(\eta)$ for $\eta \in [-1,0]$:
\begin{equation}
\begin{aligned}
&T_{21,b} = \, \frac{1}{24} \log^4 \omega -\frac{1}{6} \log^3 \omega \, \log y + \log^2 \omega \left( \frac{1}{4} \log^2y + 2 \zeta_2\right) + \\
& \hspace{1cm} + \log \omega \left( -\frac{1}{6} \log^3 y -4 \zeta_2 \log y - 8 \zeta_3 \right) \textcolor{black}{+ \frac{1}{24} \log^4 y} + \textcolor{black}{2} \zeta_2 \log^2 y + 8 \zeta_3 \log y + 9 \zeta_4\,,\\
&T_{21,c} = \, \frac{1}{24} \log^4 \omega -\frac{1}{6} \log^3 \omega \log y + \log^2 \omega \left( \frac{1}{4} \log^2 y + 2 \zeta_2 \right) + \\
& \hspace{1cm} + \log \omega \left( -\frac{1}{6} \log^3y - 4 \zeta_2 \textcolor{black}{\log y} - 10 \zeta_3 \right) \textcolor{black}{+ \frac{1}{24} \log^4 y} + 2 \zeta_2 \log^2 y + 10 \zeta_3 \log y + \frac{39}{2} \zeta_4\,,\\
& T_{22} = \frac{1}{3} \log^4 \omega - \frac{4}{3} \log^3 \omega \log y + \log^2 \omega \, (2 \log^2 y - 6 \zeta_2) + \\
&\hspace{1cm} + \log \omega \left(-\frac{4}{3} \log^3 y + 12 \zeta_2 \log y + 40 \zeta_3 \right) + \frac{1}{3} \log^4 y - 6 \zeta_2 \log^2 y \mathbin{\textcolor{black}{-}} \textcolor{black}{40 \zeta_3 \log y} - 49 \zeta_4\,.
\end{aligned}
\end{equation}
Also in this case the sum of the one-loop \eqref{eq:walking_neg_eta} and two-loop exponentiates and the logarithm of the Sudakov form factor contains at most double logarithms
\begin{equation}
\begin{aligned}
\log \mathcal{F}_2 &= -g^2 \left( \log^2 \omega - 2 \log \omega \log y + \log^2 y \right) + \\
& + g^4 \left( \zeta_2 \left( 2 \log^2 \omega - 4 \log \omega \log y + 2 \log^2 y \right) + 4 \zeta_3 (\log \omega -\log y) + 8 \zeta_4 \right)\,.
\label{eq:large_y_2loop}
\end{aligned}
\end{equation}
The two-loop expression for the walking anomalous dimension can be read off from the double logarithms 
\begin{equation}
\Gamma_{\rm walk}^{(2)}(\eta) = - 2 \zeta_2 (1 + 2 \eta + \eta^2) = -2 \zeta_2 (1 + \eta)^2\,, \quad \eta \in [-1,0]\,.
\end{equation}

\end{itemize}
In the expressions above corrections are suppressed by powers of $1/\omega$ and $y$ in \eqref{eq:small_y_2loop} or $1/y$ in \eqref{eq:large_y_2loop}. These expressions were verified by numerical studies with 
the help of the sector decomposition \cite{Binoth:2000ps,Smirnov:2008py,Kaneko:2009qx} implemented\footnote{\textcolor{black}{In fact, the subleading coefficients of $T_{22}$ for $y \ll 1$ asymptotics in Eq.\ \re{T22ySmall} were fixed by fitting numerical data. A dedicated high-precision numerical evaluation of $T_{22}$ in $D=4$, for $10^{10}\le\omega\le10^{24}$ and $10^{-10}\le y\le10^{-5}$, determines them as given in Eq.\ \re{T22ySmall}; the agreement is better than $10^{-10}$ relative at $\omega=10^{24}$, $y=10^{-10}$, with deviations decreasing like power corrections in $y$.}} in {\tt FIESTA} 
\cite{Smirnov:2021rhf}. Of course, as expected, each graph is of uniform transcendentality four. Sending $y$ to one, we observe 
that the leading and next-to-leading logarithm of $\omega$ coincide with the calculation of \cite{Henn:2011by}. The deviation 
in the other subleading contribution is expected in light of the $y \ll 1$ or $y \gg 1$ approximation implemented in the evaluation of parametric 
integrals. It would be interesting to lift these constraints in order to determine the interpolating collinear anomalous dimension.
Indeed, the single logarithms, \textcolor{black}{which take the same form $4\zeta_3(\log\omega-\log y)$ in \eqref{eq:small_y_2loop} and \eqref{eq:large_y_2loop} and vanish for $y = \omega$}, do not reduce to the 
well-known expression of the collinear anomalous dimension for $y = 1$ due to additional contributions arising from lifting the $y \ll 1$ or $y \gg 1$
approximation, which we did not account for in this work.\\

Finally, the walking anomalous dimension appearing in the Sudakov form factor takes the form
\begin{equation}
\Gamma_{\rm walk}(\eta,g) = \begin{cases} 
0 \quad \text{for} \quad \eta \leq -1\,,\\
(g^2 -2 \zeta_2 g^4) (1 + \eta)^2 + \mathcal{O}(g^6) \quad \text{for} \quad -1 \leq \eta \leq 0\,, \\
g^2(1+2 \eta-\eta^2) -  \textcolor{black}{2 g^4 \zeta_2 (1 + \eta)^2} + \mathcal{O}(g^6) \quad \text{for} \quad 0 \leq \eta \leq 1\,, \\
2 g^2 - 8 \zeta_2 g^4 + \mathcal{O}(g^6) \quad \text{for} \quad \eta \geq 1\,,
\end{cases}
\label{eq:ff_all}
\end{equation}
and it has the octagon and the cusp anomalous dimensions as boundary values
\begin{equation}
\Gamma_{\rm walk}(\eta = 0,g) 
= \frac{\Gamma_{\rm cusp}(g)}{4} \, , \qquad \Gamma_{\rm walk}(\eta \geq 1,g)  =
\frac{\Gamma_{\rm oct}(g)}{2}\, .
\label{eq:boundary_valued}
\end{equation}
Also, notice that $\Gamma_{\rm walk}(\eta)$ is a continuous function $C^0$ of $\eta$. In the following section we propose the all-loop expression for the walking anomalous dimension.

\subsection{The all-loop prediction}

In this section we make a prediction for the all-loop expression of the walking anomalous dimension using the knowledge of its boundary values and its exponentiation property. 

The latter, in particular, implies that the walking anomalous dimension can be
at most quadratic in the walking parameter $\eta$. The quadratic ansatz in the region $y < 1$ is fixed by imposing the octagon and cusp boundary values \eqref{eq:boundary_valued} to be
\begin{equation}
\Gamma_{\rm walk}(\eta,g) = \frac{\Gamma_{\rm cusp}(g)}{4} + \gamma(g) \, \eta + \left( \frac{\Gamma_{\rm oct}(g)}{2} - \frac{\Gamma_{\rm cusp}(g)}{4}-\gamma(g) \right) \eta^2\,, \quad \eta \in [0,1]\,.
\end{equation}
In particular, the ansatz cannot be fixed completely and the expression is given in terms of an unknown function $\gamma(g)$, for which the first two perturbative orders are known \eqref{eq:ff_all}:
\begin{equation}
\label{eq:gammaff2l}
\gamma(g) = 2 g^2 -  \textcolor{black}{4} \zeta_2 g^4 + \mathcal{O}(g^6)\,.
\end{equation}
At this order in the 't Hooft coupling $\gamma(g) = \ft12 \Gamma_{\rm cusp}(g) + O (g^6)$. It would be interesting to verify whether this conjectural
relation can be extended to all orders. A three-loop calculation is highly desirable.

In the other walking region $y > 1$ the ansatz is different and proportional to $(1+\eta)^2$, as suggested by the one- and two-loop result \eqref{eq:ff_all} and consistent with the fact that it should give zero at $\eta = -1$.\footnote{Strictly speaking, we could have considered an ansatz of the form $(1+\eta)(a+b \eta)$, which also goes to zero at $\eta = -1$. The cusp boundary condition fixes $a$, but the parameter $b$ would remain unfixed. The choice of $(1+\eta)^2$ is motivated and fixed by the amplitude computation of \secref{sec:amplitude}, where crossing symmetry imposes another condition. The relation \eqref{eq:satoff} allows to conclude.} It is fixed by imposing the cusp as the boundary value for $\eta = 0$:
\begin{equation}
\Gamma_{\rm walk}(\eta,g) = \frac{\Gamma_{\rm cusp}(g)}{4} (1 + \eta)^2\,, \quad \eta \in [-1,0]\,.
\end{equation}
For $\eta \geq 1$ we recover the off-shell Sudakov result \eqref{OffShellSudakov} at all-loops, since $m \leq M^2/Q$ can be neglected at double-logarithmic accuracy. Collecting the results, the all-loop prediction for the walking form factor is
\begin{equation}
\Gamma_{\rm walk}(\eta,g) = \begin{cases}
0\,, & \eta \leq -1\,, \\
\frac{\Gamma_{\rm cusp}(g)}{4} (1 + \eta)^2\,, &-1 \leq \eta \leq 0\,,\\
\frac{\Gamma_{\rm cusp}(g)}{4} + \gamma(g) \, \eta + \left( \frac{\Gamma_{\rm oct}(g)}{2} - \frac{\Gamma_{\rm cusp}(g)}{4}-\gamma(g) \right) \eta^2\,, & 0 \leq \eta \leq 1\,, \\
\frac{\Gamma_{\rm oct}(g)}{2}\,, & \eta \geq 1\,.
\end{cases}
\label{eq:ff_all_loop}
\end{equation}

This concludes our analysis of the walking anomalous dimension in the Sudakov form factor. The next section is devoted to the study of the walking behavior in the four-point scattering amplitude on the Coulomb branch of $\mathcal{N} = 4$ SYM.

\section{Walking in the four-point amplitude}
\label{sec:amplitude}

In this section, we study a four-point scattering amplitude on the Coulomb branch of $\mathcal{N}=4$ SYM in the planar limit.
We choose the vacuum expectation values (VEVs) and external states so that both the external and internal particles are massive.
We then analyze the high-energy behavior of the amplitude and show that the same “walking” phenomenon observed for the form factor also appears here, interpolating between the two asymptotic regimes identified in \cite{Caron-Huot:2021usw}, which are controlled by the cusp and octagon anomalous dimensions, respectively.

\subsection{The $(m,M)$ amplitude}

To define a four-point scattering amplitude it is sufficient to separate $k = 4$ D3-branes, as depicted in \figref{fig:setup_cb}. We consider the following color-ordered amplitude 
\begin{equation}
    i A = \langle (\Phi_I)_{N+1,N+2}(p_1) (\Phi_J)_{N+2,N+3}(p_2) (\Phi_I)_{N+3,N+4}(p_3) (\Phi_J)_{N+4,N+1}(p_4) \rangle , 
~~~ I,J \neq 8, 9 \,,
\end{equation}
with external meson states satisfying the on-shell condition
\begin{equation}
    p_i^2 = (m_i \vec{n}_i - m_{i+1} \vec{n}_{i+1})^2 \,.
\end{equation}
We generalize the case studied in \cite{Alday:2025pmg} by considering the following R-symmetry configuration\footnote{We consider the angle $\theta$ in the $(8,9)$ plane, see \cite{Flieger:2025ekn} for further details.}
\begin{equation}
    \vec{n}_1 \cdot \vec{n}_3 = \vec{n}_2 \cdot \vec{n}_4 = 1 
    \,, 
    \quad \vec{n}_i \cdot \vec{n}_{i+1} = \cos(\theta)\,.
\end{equation}
Thus opposite external legs are aligned in $R$-symmetry space, while adjacent
ones form an angle $\theta$. In this parametrization, the configuration of
\cite{Alday:2025pmg} is recovered by setting $\theta=0$.

We define the Mandelstam invariants by (we use mostly-minus signature)
\begin{equation}
    s = (p_1+p_2)^2,
    \qquad
    t = (p_2+p_3)^2 \,,
\end{equation}
and from now on we take all Coulomb-branch parameters to be equal,
\begin{equation}
    m_i \equiv m \,.
\end{equation}
Because adjacent $R$-symmetry vectors are not aligned when $\theta \neq 0$,
the external states are no longer massless. Indeed,
\begin{equation}
    p_i^2
    = 2m^2\bigl(1-\vec{n}_i \cdot \vec{n}_{i+1}\bigr)
    = 4m^2 \sin^2\!\left(\frac{\theta}{2}\right)\,.
\end{equation}
In analogy with the previous section, we consider the off-shellness to be Euclidean $M^2 \equiv -p_i^2$. This requires to analytically continue the angle $\theta \to i \vartheta$. In the calculation below this continuation is trivial and we assume that it could be equally done at finite coupling. Let us stress then that, in the following, $M^2$ has to be understood as the positive Euclidean external virtuality. \\
The generalized $R$-symmetry configuration introduces a new scale, $M$, into the problem. This extra scale gives access to new kinematic regimes
of the scattering amplitude and, in particular, makes it possible to expose
the appearance of the walking anomalous dimension.

\noindent The scattering amplitude $A$ can be conveniently written as
\begin{equation}
\begin{aligned}
    A(s,t,m,M) &= g_{YM}^2 \, \mathcal{M}(s,t,m,M,g)\,, \\
    \mathcal{M} &= 1 + g^2 \mathcal{M}^{(1)} + g^4 \mathcal{M}^{(2)}+\dots\,.
\end{aligned}
\end{equation}
In this setup, the amplitude is also a function of the mass of the external particles: the existence of this extra parameter is crucial to explore a new type of high-energy limit of the amplitude that we describe next.

\subsection{From the cusp to the octagon}

Let us first review known facts about the scattering amplitude on the Coulomb branch of $\mathcal{N} = 4$ SYM that will be useful for the discussion of the next sections.

We are interested in the high-energy limit of the amplitude as \emph{both} $s$ and $t$ are taken to be large. The simplest limit to consider is the one where we take the limit while keeping $m$ and $M$ non-zero and fixed.  
In this case the amplitude exhibits the characteristic double-logarithmic behavior \cite{Alday:2007hr,Alday:2009zm, Bruser:2018jnc}
\begin{equation}
\begin{aligned}
    \log \mathcal{M} = &-\frac{\Gamma_{\text{cusp}}(g)}{2} \log s \log t  + \mathcal{O}(\log s, \log t) \ .
\label{eq:fixed_angle}
\end{aligned}
\end{equation}

In writing \eqref{eq:fixed_angle}, it was important that $m$ and $M$ are nonzero and fixed as we take the limit. If we relax this condition, different behaviors are possible.

For example, a different double-logarithmic behavior emerges in the special case of $m=0$. In this case \cite{Caron-Huot:2021usw}
\begin{equation}
    \log \mathcal{M} = -\frac{\Gamma_{\text{oct}}(g)}{4} \log^2\left( \frac{s\, t}{M^4} \right) -\frac{1}{2} D_0(g) + \mathcal{O}(s^{-1}, t^{-1})\, \ . 
\label{eq:fixed_angle_m0}
\end{equation}

More generally, we can introduce a high-energy limit characterized by two exponents $(\eta_s, \eta_t)$, where as we take $s,t \to \infty$ we keep $s^{\eta_s}/t^{\eta_t}$ fixed. In particular, $\eta_s = \eta_t$ corresponds to the familiar fixed-angle limit. In addition, as we go to high energies, we can scale the ratio of the internal mass $m$, and the external mass $M$. To characterize different behaviors we find it convenient to introduce the following dimensionless variables
\be
\omega_s \equiv \frac{-s}{M^2}\,, \quad \omega_t \equiv \frac{-t}{M^2}\,, \quad y = \frac{m^2}{M^2} \ . 
\label{eq:definition_omegas_v2}
\ee
In terms of these dimensionless variables we consider a family of high-energy limits $\omega_s, \omega_t \to \infty$ with
\be
\eta_s = -\frac{\log y}{\log \omega_s}\,, \quad \eta_t = -\frac{\log y}{\log \omega_t} \ , 
\label{eq:double_scaling_st}
\ee
kept fixed. 

In the next section, we will show that the walking anomalous dimension behavior found in the Sudakov form factor that interpolates between \eqref{eq:fixed_angle} and \eqref{eq:fixed_angle_m0} generalizes to the $(m,M)$ scattering amplitude on the Coulomb branch of $\mathcal{N} = 4$ SYM. We will present evidence that
\begin{equation}
\begin{aligned}
    &\log \mathcal{M} = -\Gamma_{\text{walk}}(\eta_s,\eta_t,g) \log\omega_s \log\omega_t + \mathcal{O}(\log\omega_s,\log\omega_t)\, .
\label{eq:final_res_walking}
\end{aligned}
\end{equation}
with $\Gamma_{\text{walk}}(\eta_s,\eta_t,g)$ the walking anomalous dimension that interpolates between the cusp (as $\eta_s, \eta_t \to 0$), and the octagon anomalous dimensions (as $\eta_s, \eta_t \to 1$). 

\subsection{A one-loop derivation}

\begin{figure}[h!]
    \centering
    \includegraphics[width=0.4\linewidth]{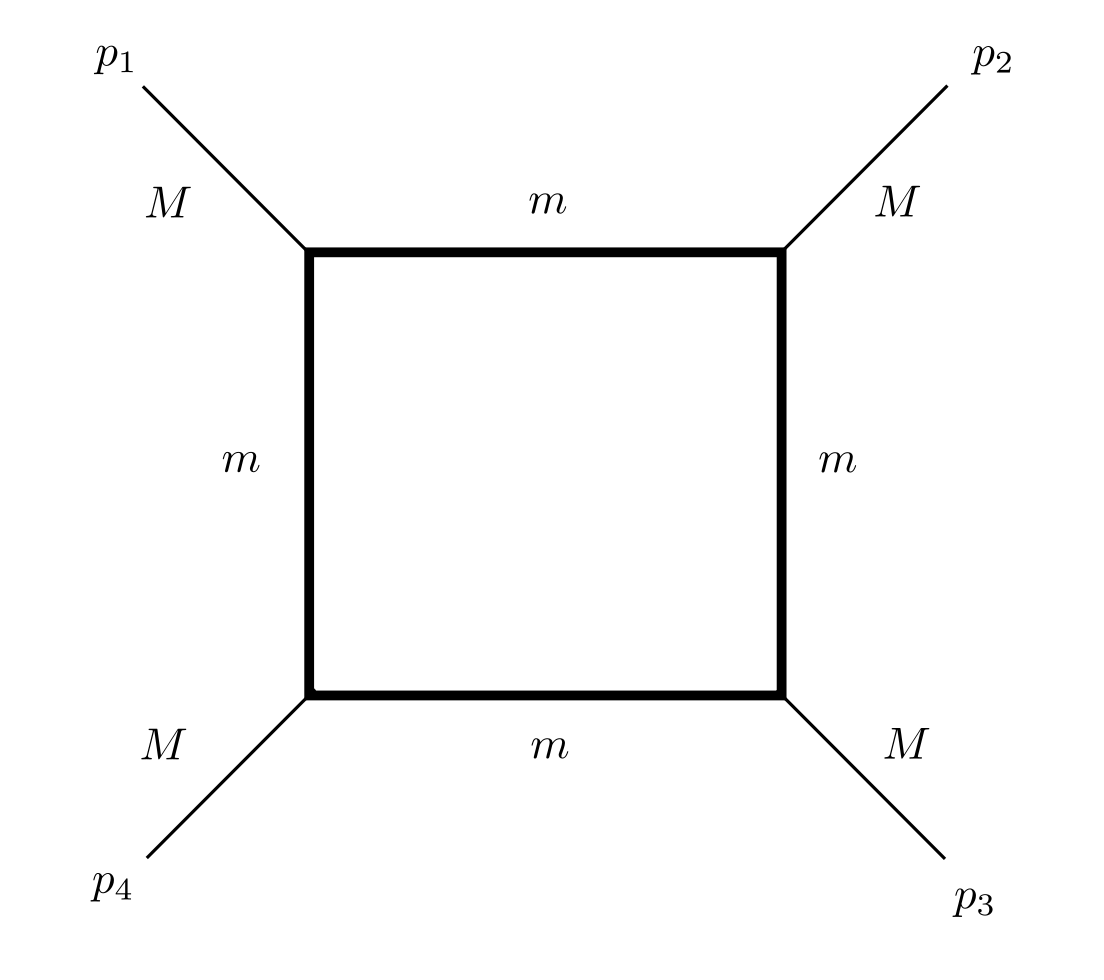}
    \caption{The one-loop box with massive external particles $p_i^2 = -M^2$ and massive W-bosons of mass $m$ in the internal loop.}
    \label{fig:one_loop_mM}
\end{figure}

In this section we will repeat the analysis done in \secref{sec:one_loop_ff_1} for the one-loop amplitude on the Coulomb branch represented in \figref{fig:one_loop_mM}. We will derive an expression for the one-loop walking anomalous dimension consistent with \eqref{eq:satoff} and \eqref{eq:ff_one_loop_all} and we will show that \eqref{eq:final_res_walking} holds at one-loop.

In order to derive the expression of $\Gamma_{\text{walk}}^{(1)}(\eta_s,\eta_t)$ it is convenient to use the dispersion representation of the one-loop box derived by Mandelstam in \cite{Mandelstam:1959bc}:
\begin{equation}
    \mathcal{M}^{(1)} = -2 \, st\,\int_{4m^2}^{\infty} \frac{ds'dt'}{(s'-s)(t'-t)} \frac{\theta(s't' - 4m^2(s'+t'+4M^2)-4M^4)}{\sqrt{s't'(s't' - 4m^2(s'+t'+4M^2)-4M^4)}}\,.
\end{equation}
After performing the integral in $t'$ it is possible to derive the standard dispersion representation\footnote{The dispersion representation \eqref{eq:dispersion_rep_one_loop} is valid for $p_i^2 \leq 2m^2$. For $p_i^2 > 2m^2$ there is an additional contribution coming from the anomalous threshold, see e.g. \cite{Correia:2022dcu} for a recent discussion. Since we are working in the Euclidean regime $p_i^2 = -M^2 < 0$, this is not relevant for our analysis.}
\begin{equation}
\begin{aligned}
 &\mathcal{M}^{(1)}=  s t\, \int_{4m^2}^\infty ds' \hspace{1mm}{\mathcal{M}_{s}(s',t) \over s'-s}\,, \\
 &\mathcal{M}_s(s,t) = \frac{4 \sinh ^{-1}\left(\frac{1}{2} \sqrt{-\frac{t \left(s-4 m^2\right)}{m^2 \left(s+4
   M^2\right)+M^4}}\right)}{\sqrt{-t} \sqrt{s \left(4 m^2 \left(+4 M^2+s+t\right)+4
   M^4-s t\right)}} \ . 
\label{eq:dispersion_rep_one_loop}
\end{aligned}
\end{equation}

In the following we will derive the expression for $\Gamma^{(1)}_{\text{walk}}(\eta_s,\eta_t)$ in the simple case $\eta_s = \eta_t$ and then we will discuss the most general one $\eta_s \neq \eta_t$. 

\subsubsection{The case $s = t$}

We proceed as in \secref{AnticipSection} by doing a convenient change of variables as in \eqref{eq:definition_omegas_v2} with
\begin{equation}
    \omega_s = \omega_t \equiv \omega\,.
\end{equation}
In this case the expansions \eqref{eq:fixed_angle} and \eqref{eq:fixed_angle_m0} have the same leading-logarithmic behavior 
\begin{equation}
\begin{aligned}
&\log \mathcal{M}\Big{|}_{(s,t) \to \infty,\, m, M-\text{fixed}} = -\frac{\Gamma_{\text{cusp}}(g)}{2} \log^2\left(\omega\right) + \mathcal{O}(\log \omega)\,, \\ 
&\log \mathcal{M}\Big{|}_{(s,t) \to \infty, m = 0} = -\Gamma_{\text{oct}}(g) \log^2\left(\omega\right) + \mathcal{O}(\omega^0)\,,
\label{eq:fixed_angle_all}
\end{aligned}
\end{equation}
and the walking anomalous dimension $\Gamma_{\text{walk}}(\eta,g)$ interpolates between the two anomalous dimensions that appear in the different regimes.

\begin{figure}[h!]
    \centering
    \includegraphics[width=0.85\linewidth]{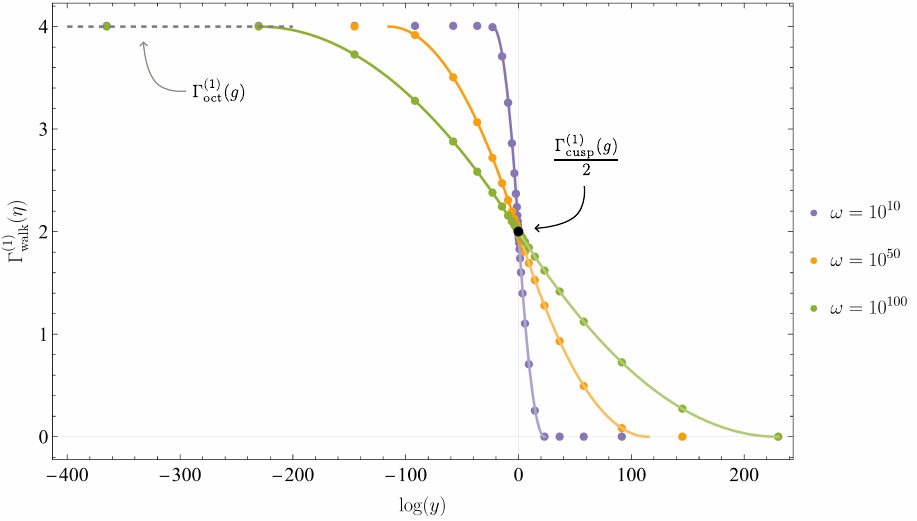}
    \caption{The one-loop walking anomalous dimension $\Gamma_{\text{walk}}^{(1)}(\eta) = \lim_{\omega \to \infty} (-\mathcal{M}^{(1)}/\log^2 \omega)$ obtained from the one-loop amplitude at $s = t$ \eqref{eq:disp_one_loop} as a function of $\log y \equiv \log (m^2/M^2)$ and different, large, values of $\omega$. The dots are obtained from the numerical evaluation of \eqref{eq:disp_one_loop}, while the curves are given by the explicit analytic expressions \eqref{eq:inter_cusp_amplitude}. We observe complete matching between the two for high enough energies.}
    \label{fig:walking_amp}
\end{figure}

In the variables defined above the one-loop dispersion representation of the amplitude takes the form
\begin{equation}
    \mathcal{M}^{(1)} = -4\, \omega^{3/2}\,\int_0^{\infty} dx\, \frac{\sinh^{-1}\left(\frac{\sqrt{x \omega}}{\sqrt{1+4y(x+y+1)}}\right)}{(\omega + 4(x+y))\sqrt{x+y}\sqrt{1+ x \omega + 4 y (x + y + 1)}}\,,
\label{eq:disp_one_loop}
\end{equation}
where we recall that 
\begin{equation}
x= \frac{s'}{4M^2}-y\,, \quad y = \frac{m^2}{M^2}\,.
\end{equation}
We are interested in the double-logarithmic behavior of the expression above in the high-energy limit
\begin{equation}
\omega \to \infty\,, \quad \eta = -\frac{\log y}{\log \omega} \quad\text{fixed}\,.
\label{eq:double_scaling_limit}
\end{equation}
First we proceed with a numerical analysis of \eqref{eq:disp_one_loop}. As done for the form factor, we numerically evaluate the dispersion representation for large values of $\omega$ and plot $(-\mathcal{M}^{(1)}/\log^2 \omega)$ as a function of $\log(y)$, which goes to $\Gamma^{(1)}_{\rm walk}(\eta)$ as $\omega \to \infty$. The result in \figref{fig:walking_amp} suggests the same walking behavior observed in the Sudakov form factor as shown in \figref{fig:walking_ff}. 

\begin{figure}[h!]
    \centering
    \includegraphics[width=0.85\linewidth]{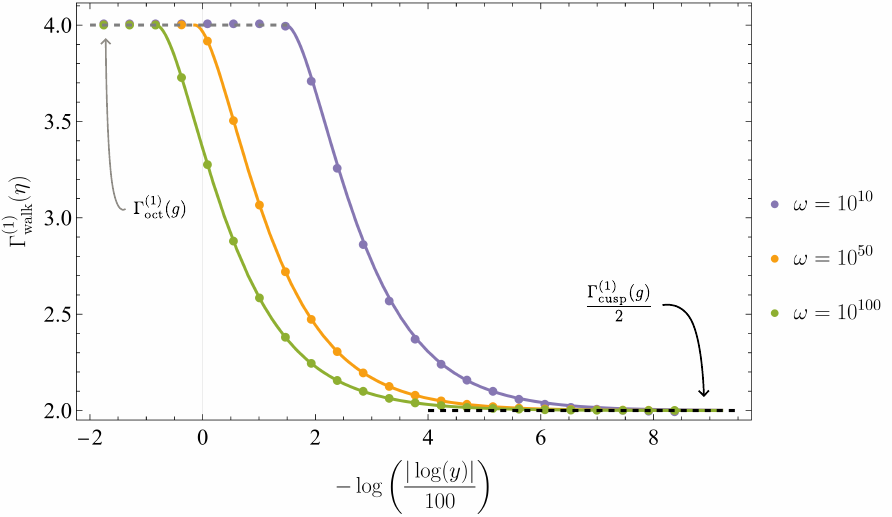}
    \caption{Zoom of the octagon-to-cusp-transition region ($y < 1$) for one-loop walking anomalous dimension $\Gamma_{\text{walk}}^{(1)}(\omega,y) = \lim_{\omega \to \infty} (-\mathcal{M}^{(1)}/\log^2 \omega)$. The plot is obtained by numerically evaluating \eqref{eq:disp_one_loop} as a function of $-\log\left(\frac{|\log y|}{100}\right) \equiv -\log\left(\frac{|\log (m^2/M^2)|}{100}\right)$ at different, large, values of $\omega$. The curve is the analytic expression \eqref{eq:inter_cusp_amplitude}.}
    \label{fig:walking_amp_zoom}
\end{figure}

The octagon-to-cusp transition occurs for $y < 1$, with the same octagon shoulder forming for $y \leq 1/\omega$. A zoom on this specific region is presented in \figref{fig:walking_amp_zoom}. For $y > 1$, the walking anomalous dimension interpolates between the cusp and $0$, with another shoulder forming for $y \geq \omega$. In the following we will derive analytically the expression of the walking anomalous dimension at one-loop by analyzing the dispersion relation of the amplitude. In particular, we identify the regions of the integral that give double logarithms, as presented in detail in \appref{sec:detailed_analysis_amplitude}.

First, we focus on the case $y < 1$. To analyze analytically the octagon-to-cusp transition happening for $y \in [1/\omega,1]$, it is enough to consider the small-$y$ expansion of \eqref{eq:disp_one_loop}, in the high-energy limit \eqref{eq:double_scaling_limit}. It is important to remark that the result is not obtained by setting $y = 0$: the $y$ dependence that gives the logarithmic behavior has to be kept, in particular the integral probes the scale $x \sim y^{-1}$ and $x \sim y$.

\noindent At the leading double-logarithmic accuracy we obtain
\begin{equation}
    \mathcal{M}^{(1)} \Big{|}_{y \ll 1} = -2\, \int_0^{\infty} dx\, \frac{\log\left(\frac{1+2 x \omega + 2 \sqrt{x \omega(1+x \omega)}}{1+4(x+1)y}\right)}{(1 + 4 x/\omega)\sqrt{x+y}\sqrt{(x+1/\omega)}}\,,
\end{equation}
where we have also expressed the $\sinh^{-1}$ as a $\log$. As suggested by the numerical analysis \figref{fig:walking_amp_zoom}, there are two different regions in which $\Gamma_{\text{walk}}^{(1)}(\eta)$ behaves differently for $y < 1$:
\begin{equation}
\begin{aligned}
&y < \omega^{-1} < 1 \quad \Rightarrow \quad \frac{M^2}{m^2} > \frac{-s}{M^2},\frac{-t}{M^2} > 1\,, \\
&1 > y > \omega^{-1} \quad \Rightarrow \quad  \frac{-s}{M^2}, \frac{-t}{M^2} > \frac{M^2}{m^2} > 1 \,.
\end{aligned}
\end{equation}

From the first region, in the limit $\omega = -s/M^2, -t/M^2 \gg1$,  we get the following hierarchy of scales 
\begin{equation}
\frac{M^4}{m^2} > -s,-t \gg M^2\,.
\end{equation}
The scattering energy is much larger than $M^2$ but still not large enough to see the effect of having a finite mass $m$, which is then effectively zero. In this regime we see the same behavior as in \eqref{eq:fixed_angle_m0}, with the octagon anomalous dimension appearing as the double-logarithmic coefficient. In the second region, for $1 > y > 1/\omega$ the hierarchy of scales is different
\begin{equation}
-s,-t > \frac{M^4}{m^2} > M^2\,.
\end{equation}
In particular, the scattering energy is large enough to resolve $m$ and the result is sensitive to the finite internal mass $m$. In this regime we will eventually get \eqref{eq:fixed_angle} with the result controlled by $\Gamma_{\text{cusp}}(g)$.

The behavior of $\mathcal{M}^{(1)}$ in the octagon shoulder $y < \omega^{-1}$, corresponding to $\eta > 1$ is given by
\begin{equation}
\begin{aligned}
    \mathcal{M}^{(1)}\Big{|}_{y < \omega^{-1}} &= -2\, \int_{\omega^{-1}}^{\omega} dx\, \frac{\log\left(\frac{4 x \omega}{1+4 y}\right)}{x} + \mathcal{O}(\log \omega) =\\ 
     &= -4 \log^2\omega + \mathcal{O}(\log\omega) + \mathcal{O}(y) = \\
     &= -\Gamma^{(1)}_{\text{walk}}(\eta) \log^2\omega + \mathcal{O}(\log\omega) + \mathcal{O}(y)\,, \quad \eta > 1\,,
\end{aligned}
\end{equation}
while in the first walking region $y > \omega^{-1}$, equivalent to $\eta \in [0,1]$
\begin{equation}
\begin{aligned}
    \mathcal{M}^{(1)} \Big{|}_{y > \omega^{-1}} &= 
    -2\, \int_{y}^{y^{-1}} dx\, \frac{\log\left(\frac{4 x \omega}{1+4 y}\right)}{x} -2\, \int_{y^{-1}}^{\omega} dx\, \frac{\log\left(\frac{4 x \omega}{4 x y}\right)}{x} + \mathcal{O}(\log y, \log\omega) = \\
    &= -\left(2 + 4 \eta -2 \eta^2 \right) \log^2\omega + \mathcal{O}(\log y, \log \omega) = \\
    &= -\Gamma^{(1)}_{\text{walk}}(\eta) \log^2\omega + \mathcal{O}(\log y, \log \omega)\,, \quad \eta \in [0,1]\,.
\end{aligned}
\label{eq:gamma_2}
\end{equation}
As anticipated, the above expression for $\Gamma^{(1)}_{\text{walk}}(\eta) = 2 g^2 (1+ 2 \eta - \eta^2)$, valid for $\eta \in [0,1]$, has the one-loop octagon and one-loop cusp anomalous dimensions as boundary values
\begin{equation}
\Gamma^{(1)}_{\rm walk}(0) 
= \frac{\Gamma^{(1)}_{\rm cusp}}{2} \, , \qquad \Gamma^{(1)}_{\rm walk}(1) =\Gamma^{(1)}_{\rm oct}\, .
\end{equation}

We focus now on the case $y > 1$, where the walking anomalous dimension interpolates between the cusp and zero. In order to find its analytic expression one has to consider the opposite $y \gg 1$ limit in the integrand of \eqref{eq:disp_one_loop}
\begin{equation}
\mathcal{M}^{(1)}\Big{|}_{y \gg 1} = -2 \int_0^{\infty} dx \, \frac{\log \left( \frac{(\sqrt{x \omega} + \sqrt{x \omega + 4 y (x+y)})^2}{4 y (x+y)} \right)}{(1 + 4 (x+y)/\omega) \sqrt{x+y} \sqrt{(x + 4y (x+y)/\omega)}}\,,
\end{equation}
where we have expressed the $\sinh^{-1}$ as a $\log$.
In the interpolating region $1 < y < \omega$, corresponding to $\eta \in [-1,0]$, the relevant contribution is
\begin{equation}
\begin{aligned}
\mathcal{M}^{(1)}\Big{|}_{ y < x < \omega} &= -2 \int_{y}^{\omega} dx \, \frac{\log\left(\frac{(\sqrt{x \omega} + \sqrt{x \omega + 4 x y})^2}{4 x y}\right)}{\sqrt{x (x + 4 x y/\omega)}} = -2 \log^2\left( \frac{\omega}{y}\right) + \mathcal{O}(y^{-1}) = \\
&= -2 (1 + \eta)^2 \log^2 \omega + \mathcal{O}(y^{-1}) = \\
&= -\Gamma_{\rm walk}^{(1)}(\eta) \log^2 \omega + \mathcal{O}(y^{-1})\,, \quad \eta \in [-1,0]\,.
\end{aligned}
\label{eq:neg_eta_one_loop}
\end{equation}
For the other shoulder $y > \omega$, $\Gamma_{\text{walk}}^{(1)}(y > \omega) = 0$ since in the limit $m^2 \gg -s,-t$ the loop degrees of freedom are not dynamical.\\

Combining all the expressions found above, the one-loop walking anomalous dimension appearing in the one-loop amplitude for $s = t$ takes the form
\begin{equation}
    \Gamma_{\text{walk}}^{(1)}(\eta) = g^2 \begin{cases}
    0 \quad \text{for} \quad \eta \leq -1\,, \\
    2 (1 + \eta)^2 \quad \text{for} \quad -1 \leq \eta \leq 0\,,\\
    2 (1 + 2 \eta -\eta^2) \quad \text{for} \quad 0 \leq \eta \leq 1\,, \\
    4 \quad \text{for} \quad \eta \geq 1\,,
    \end{cases}
\label{eq:inter_cusp_amplitude}
\end{equation}
in agreement with the relation \eqref{eq:satoff} to the form factor anomalous dimension \eqref{Gwalk1}. The function is $C^1$. As anticipated, the walking anomalous dimension interpolates between the cusp and the octagon as in \eqref{eq:fixed_angle_all} and in particular
\begin{equation}
\Gamma_{\text{walk}}^{(1)}(\eta = 0) = 2g^2 = \frac{\Gamma_{\text{cusp}}^{(1)}}{2}\,, \quad \Gamma_{\text{walk}}^{(1)}(\eta \geq 1) = 4g^2 = \Gamma_{\text{oct}}^{(1)}\,.
\end{equation}

\noindent In the next section we analyze the case of $\omega_s \neq \omega_t$.

\subsubsection{Case $s \neq t$}

We consider the most general case $s \neq t$. From  \eqref{eq:fixed_angle} and \eqref{eq:fixed_angle_m0} we expect to have a different logarithmic behavior in the different regimes:
\begin{equation}
\begin{aligned}
&\log \mathcal{M}\Big{|}_{m, M - \text{fixed}} = -\frac{\Gamma_{\text{cusp}}(g)}{2} \log\omega_s\log\omega_t + \mathcal{O}(\log\omega_s,\log\omega_t)\,, \\
&\log \mathcal{M}\Big{|}_{m=0, M \to 0} = - \frac{\Gamma_{\text{oct}}(g)}{4} \left( \log^2 \omega_s + \log^2 \omega_t + 2 \log \omega_s \log \omega_t \right) + \mathcal{O}(s^0,t^0) = \\
&\hspace{2.8cm}= - \frac{\Gamma_{\text{oct}}(g)}{2} \left( 1 + \frac{\log \omega_s}{2 \log\omega_t}+\frac{\log\omega_t}{2 \log \omega_s}  \right) \log \omega_s \log \omega_t + \mathcal{O}(s^0,t^0)\,.
\end{aligned}
\label{eq:octagon_region}
\end{equation}

In the following, we explicitly observe this behavior and we derive the one-loop walking anomalous dimension that interpolates between the two regimes
\begin{equation}
\Gamma_{\text{walk}}^{(1)}(\eta_s,\eta_t) \equiv \lim_{\omega_s, \omega_t \to \infty}-\frac{\mathcal{M}^{(1)}}{\log\omega_s \log \omega_t}\,,
\end{equation}
since $\log \mathcal{M} = \log(1+ g^2 \mathcal{M}^{(1)} + \mathcal{O}(g^4)) = g^2 \mathcal{M}^{(1)} + \mathcal{O}(g^4)$.
By using the change of variables \eqref{eq:definition_omegas_v2}, the one-loop dispersion representation can be written as
\begin{equation}
    \mathcal{M}^{(1)} = -4 \omega_s \int_0^{\infty} dx \, \frac{\sinh^{-1}\left(\frac{\sqrt{x \omega_t}}{\sqrt{1+4y(x+y+1)}}\right)}{(\omega_s + 4 (x+y)) \sqrt{\frac{x+y}{\omega_t}}\sqrt{1+x \omega_t + 4 y(x+y+1)}}\,.
\label{eq:disp_rel_different_st}
\end{equation}
Since $\mathcal{M}^{(1)}$ is crossing symmetric in $s,t$ we can focus on the case $\eta_t < \eta_s$. The other $\eta_t > \eta_s$ follows by crossing. As in the previous section, we are interested in the leading logarithmic contribution to the one-loop amplitude in the high-energy limit \eqref{eq:double_scaling_st}.

\begin{figure}[h!]
    \centering
    \includegraphics[width=0.85\linewidth]{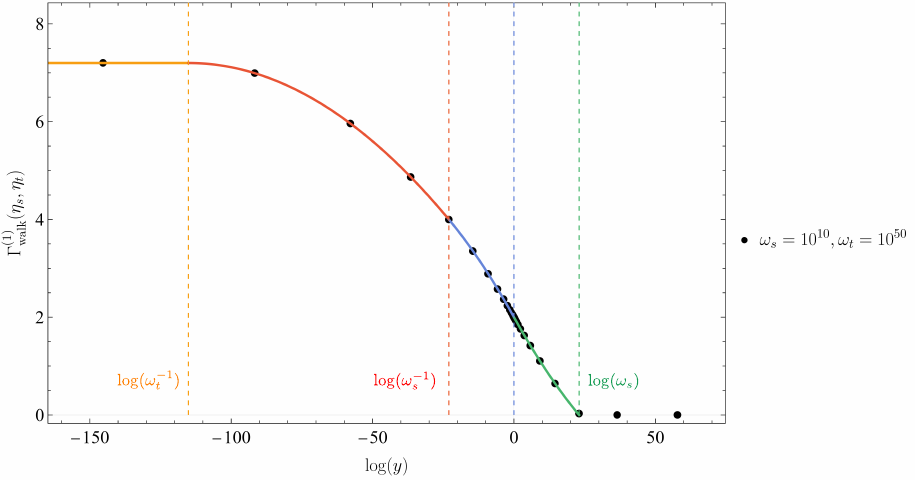}
    \caption{The one-loop walking anomalous dimension $\Gamma_{\text{walk}}^{(1)}(\eta_s, \eta_t) = \lim_{\omega_s,\omega_t \to \infty} (-\mathcal{M}^{(1)}/(\log \omega_s \log \omega_t))$ obtained from the one-loop amplitude at $s \neq t$ \eqref{eq:disp_rel_different_st} as a function of $\log y \equiv \log (m^2/M^2)$ and different, large, values of $\omega_s, \omega_t$. The dots are obtained from the numerical evaluation of \eqref{eq:disp_rel_different_st}, while the curves are given by the explicit analytic expressions \eqref{eq:Gamma-walk-etas-ordered}. We observe the emergence of five different regions, with a different behavior for $\Gamma^{(1)}_{\rm walk}(\eta_s,\eta_t)$.}
\label{fig:walking_amp_diff_st_new}
\end{figure}

First, we proceed with a numerical evaluation of \eqref{eq:disp_rel_different_st} in the high-energy limit of interest and we plot $(-\mathcal{M}^{(1)}/(\log \omega_s \log \omega_t))$ as a function of $\log(y)$, which goes to $\Gamma^{(1)}_{\rm walk}(\eta_s,\eta_t)$ as $\omega_s,\omega_t \to \infty$. The result is presented in \figref{fig:walking_amp_diff_st_new}. 

\begin{figure}[h!]
    \centering
    \includegraphics[width=0.85\linewidth]{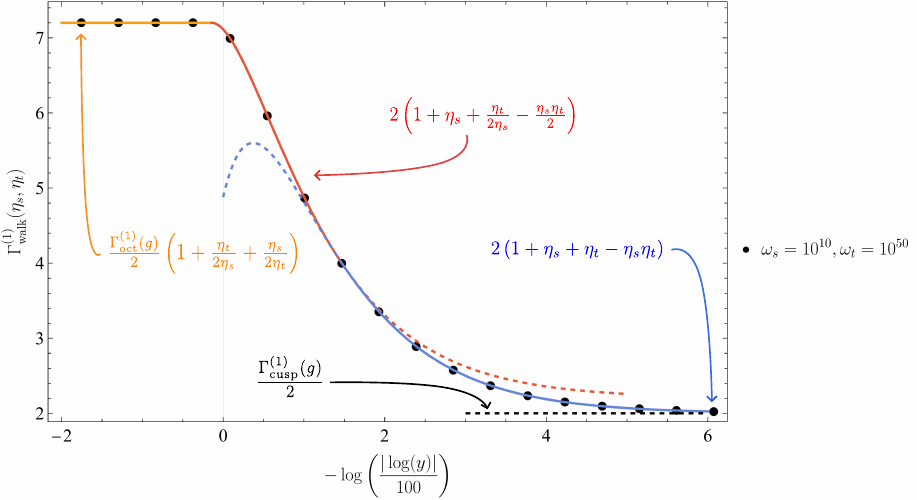}
    \caption{The one-loop walking anomalous dimension $\Gamma_{\text{walk}}^{(1)}(\eta_s,\eta_t) = \lim_{\omega_s,\omega_t \to \infty} (-\mathcal{M}^{(1)}/(\log\omega_s \log \omega_t))$ obtained from the one-loop amplitude at $s \neq t$ \eqref{eq:disp_rel_different_st} as a function of $\log y \equiv \log (m^2/M^2)$ and different, large, values of $\omega_s, \omega_t$ with $\omega_s < \omega_t$. The dots are obtained from the numerical evaluation of \eqref{eq:disp_rel_different_st}, while the curves are given by the explicit analytic expressions \eqref{eq:Gamma-walk-etas-ordered} in the different regions with $\eta_s, \eta_t \geq 0$. The dashed blue line is a continuation of the expression \eqref{eq:gamma_ybig} for $y < \omega_s^{-1}$ and the dashed red line is the continuation of \eqref{eq:intermediate_y} for $y > \omega_s^{-1}$. This clearly shows that three different functions are needed to reproduce the result, while for the $s = t$ case only two were needed.}
\label{fig:gamma_different_st_zoom}
\end{figure}

By zooming in the octagon-to-cusp transition region, happening for $y < 1$, it is possible to notice the emergence of three regions in which $\Gamma_{\text{walk}}^{(1)}(\eta_s,\eta_t)$ behaves differently: 
\begin{equation}
    y < \omega_t^{-1}\,, \quad  \omega_t^{-1} < y < \omega_s^{-1}\,, \quad  y > \omega_s^{-1}\,,
\end{equation}
as shown in \figref{fig:gamma_different_st_zoom}.

For now we focus on the derivation of the analytic expression for the walking anomalous dimension in the region $y < 1$, given in \figref{fig:gamma_different_st_zoom}.
As done in the previous sections, it is possible to simplify the dispersion representation for the one-loop box by considering $y$ to be small in the limit \eqref{eq:double_scaling_st}:
\begin{equation}
    \mathcal{M}^{(1)}\Big{|}_{y \ll 1} = -2 \, \int_0^{\infty} dx\, \frac{\log\left(\frac{1+2 x \omega_t + 2 \sqrt{x \omega_t(1+x \omega_t)}}{1+4(x+1)y}\right)}{(1 + 4 x/\omega_s)\sqrt{x+y}\sqrt{(x+1/\omega_t)}}\,.
\label{eq:one_loop_walk_different_st}
\end{equation}
In particular, the $y$ dependence that gives the logarithmic behavior is kept: the integral probes the scales $x \sim y$ and $x \sim y^{-1}$.

In the following we present the result for the case $y < 1$ and highlight the three different cases. 
\noindent First, for $y > \omega_s^{-1}$ the double logarithms originate from the following integration regions
\begin{equation}
    \begin{aligned}
        \mathcal{M}^{(1)} &= -2 \Big( \int_{y}^{y^{-1}} dx\, \frac{\log\left(\frac{4 x \omega_t}{1+4y}\right)}{x} + \int_{y^{-1}}^{\omega_s} dx\, \frac{\log\left(\frac{\omega_t}{y}\right)}{x} \Big) + \mathcal{O}(\log\omega_s, \log \omega_t, \log y) = \\
        &= -2 \left(1-\frac{\log y}{\log \omega_s}-\frac{\log y}{\log \omega_t}-\frac{\log^2y}{\log \omega_s \log \omega_t}\right) \log \omega_s \log\omega_t + \mathcal{O}(\log\omega_s, \log \omega_t, \log y) = \\
        &= -\Gamma^{(1)}_{\text{walk}}(\eta_s,\eta_t) \log \omega_s \log\omega_t + \mathcal{O}(\log\omega_s, \log \omega_t, \log y)\,.
    \end{aligned}
\label{eq:gamma_ybig} 
\end{equation}

\noindent On the other hand, for $\omega_t^{-1} < y < \omega_s^{-1}$ 
\begin{equation}
    \begin{aligned}
        \mathcal{M}^{(1)} &= -2 \int_{y}^{\omega_s} dx\, \frac{\log\left(\frac{4 x \omega_t}{1+4y}\right)}{x} + \mathcal{O}(\log\omega_s, \log \omega_t, \log y) = \\
        &= -2 \left(1-\frac{\log y}{\log \omega_s}-\frac{\log^2y}{2\log \omega_s \log \omega_t} + \frac{\log \omega_s}{2 \log \omega_t}\right) \log \omega_s \log\omega_t + \mathcal{O}(\log\omega_s, \log \omega_t, \log y) = \\
        &= -\Gamma^{(1)}_{\text{walk}}(\eta_s,\eta_t) \log \omega_s \log\omega_t + \mathcal{O}(\log\omega_s, \log \omega_t, \log y)\,.
    \end{aligned}
\label{eq:intermediate_y}
\end{equation}

\noindent In the last region $y < \omega_t^{-1}$, characterized by the octagon-shoulder, the relevant double logarithms are given by 
\begin{equation}
    \begin{aligned}
        \mathcal{M}^{(1)} &= -2 \int_{\omega_t^{-1}}^{\omega_s} dx\, \frac{\log\left(\frac{4 x \omega_t}{1+4y}\right)}{x} + \mathcal{O}(\log\omega_s, \log \omega_t, \log y) = \\
        &= -2 \left( 1 + \frac{\log \omega_s}{2 \log\omega_t}+\frac{\log\omega_t}{2 \log \omega_s}\right) \log \omega_s \log\omega_t + \mathcal{O}(\log\omega_s, \log \omega_t, \log y) = \\
        &= -\Gamma^{(1)}_{\text{walk}}(\eta_s,\eta_t) \log \omega_s \log\omega_t + \mathcal{O}(\log\omega_s, \log \omega_t, \log y)\,.
    \end{aligned}
\label{eq:gamma_ysmall}
\end{equation}

This concludes the analysis for $y  < 1$. The other region $y > 1$ can be easily obtained by recalling the result in the $s = t$ case \eqref{eq:inter_cusp_amplitude}. There, the only nonzero contribution is given by $\Gamma_{\rm walk}^{(1)}(\eta) = 2(1+\eta)^2$ for $\eta \in [-1,0]$. By crossing, the generalization to the $s \neq t$ case is simply
\begin{equation}
\Gamma_{\rm walk}^{(1)}(\eta_s,\eta_t) = 2 (1 + \eta_s) (1+ \eta_t)\,, \quad -1 \leq \eta_t \leq \eta_s \leq 0\,,
\end{equation}
which also matches the numerical evaluation of \figref{fig:walking_amp_diff_st_new}.

We can finally combine the results for the one-loop walking anomalous dimension in the $s \neq t$, $\eta_t < \eta_s$ case as follows
\begin{equation}
\Gamma_{\rm walk}^{(1)}(\eta_s,\eta_t)
=
2g^2
\begin{cases}
0, & \eta_t \leq \eta_s \leq 0,\,\, \eta_t \leq -1, \\
(1+\eta_s) (1 + \eta_t), & -1 \leq \eta_t \leq \eta_s \leq 0,\\
1+\eta_s+\eta_t-\eta_s\eta_t,
& 0 \leq \eta_t \leq \eta_s \leq 1, \\[4pt]
1+\eta_s+\dfrac{\eta_t}{2\eta_s}-\dfrac{\eta_s\eta_t}{2},
& 0 \leq \eta_t \leq 1 \leq \eta_s, \\[8pt]
1+\dfrac{\eta_t}{2\eta_s}+\dfrac{\eta_s}{2\eta_t},
& 1 \leq \eta_t \leq \eta_s .
\end{cases}
\label{eq:Gamma-walk-etas-ordered}
\end{equation}
The complementary region $\eta_s<\eta_t$ is obtained by crossing. The function $\Gamma^{(1)}_{\text{walk}}(\eta_s,\eta_t)$ is $C^1$.
As anticipated, in the limit $\eta_s, \eta_t \to 0$ the value of $\Gamma_{\text{cusp}}^{(1)}/2 = 2 g^2$ is recovered, while for $1 \leq \eta_t \leq \eta_s$ we reproduce the behavior of \eqref{eq:fixed_angle_m0}, with $\Gamma_{\text{oct}}^{(1)} = 4 g^2$, as expected.\\

In the following section we will make a prediction for the all-loop expression of the walking anomalous dimension appearing in the four-point amplitude for $s \neq t$, in a similar way as done for the walking Sudakov form factor.

\subsection{The all-loop prediction}

In this section we work out the expression for the walking anomalous dimension appearing in the amplitude in the most general case $\eta_s \neq \eta_t$ and at finite coupling $g$.
First, we make the following ansatz for $\eta_s, \eta_t \geq 0$
\begin{equation}
\Gamma_{\text{ansatz}}(\eta_s,\eta_t,g) = \Tilde{\gamma} + \left( a_1 \frac{\eta_t}{\eta_s} + a_2 \frac{\eta_s}{\eta_t} \right) + \left( b_1 \eta_s + b_2 \eta_t \right) + c \, \eta_s \eta_t \,,
\label{eq:all_ansatz}
\end{equation}
based on crossing symmetry and the fact that we expect at most double-logarithmic asymptotic behavior of the amplitude. We will focus on the case $\eta_t < \eta_s$, the complementary region can be obtained by crossing.
\subsubsection*{Region $0 \leq \eta_t \leq \eta_s \leq 1$}
In this region we have to uplift the predicted all-loop expression for the $\eta_s = \eta_t$ case of the Sudakov form factor
\begin{equation}
\Gamma_{\rm walk}(\eta,\eta,g)
= 2 \left(\frac{\Gamma_{\rm cusp}}{4} + \gamma \, \eta + \left(\frac{\Gamma_{\rm oct}}{2} - \frac{\Gamma_{\rm cusp}}{4} - \gamma \right) \eta^2\right)\, ,
\end{equation}
to the $\eta_s \neq \eta_t$ case
\begin{equation}
\Gamma_{\rm walk}(\eta_s,\eta_t,g)
= 2 \left(\frac{\Gamma_{\rm cusp}}{4} + \frac{\gamma}{2} \, (\eta_s + \eta_t) + \left(\frac{\Gamma_{\rm oct}}{2} - \frac{\Gamma_{\rm cusp}}{4} - \gamma \right) \eta_s \eta_t\right)\,.
\label{eq:all_loop_sudakov_different_st_intermediate}
\end{equation}
In doing the uplift, we assumed that terms of the form $\eta_s/\eta_t$ do not appear based on the expected double-logarithmic behavior in the limit $\eta_s, \eta_t \to 0$. 
\subsubsection*{Region $1 \leq \eta_t \leq \eta_s$}
By matching $\Gamma_{\text{ansatz}}(\eta_s, \eta_t = 1,g)$ to the behavior in the octagon region \eqref{eq:octagon_region} for $\eta_t = 1$ we get
\begin{equation}
\begin{aligned}
\Gamma_{\text{ansatz}}(\eta_s,\eta_t = 1,g) &= (\Tilde{\gamma} + b_2) +  \frac{a_1}{\eta_s} +  (a_2 + b_1 + c) \eta_s \equiv \frac{\Gamma_{\text{oct}}}{2} \left( 1 + \frac{1}{2 \eta_s} + \frac{\eta_s}{2} \right)\,,
\end{aligned}
\end{equation}
which fixes three of the free parameters. 
\subsubsection*{Region $0 \leq \eta_t \leq 1 \leq \eta_s$}
Finally, in the intermediate region $\Gamma_{\text{ansatz}}(\eta_s = 1,\eta_t,g)$ can be matched with $\Gamma_{\text{walk}}(\eta_s = 1,\eta_t,g)$. This fixes two of the remaining free parameters, leaving two of them $\gamma, \Tilde{\gamma}$ unfixed.
The ansatz is then fixed to be
\begin{equation}
\begin{aligned}
\Gamma_{\text{ansatz}}(\eta_s,\eta_t,g) = \left(\frac{\Gamma_{\text{cusp}}}{2} + \gamma \right) \eta_s (1-\eta_t) + \frac{\Gamma_{\text{oct}}}{4} \frac{\eta_t}{\eta_s} (1+\eta_s)^2 + \Tilde{\gamma}(\eta_s-1)(\eta_t-1)\,.
\end{aligned}
\end{equation}
\subsubsection*{Region $-1 \leq \eta_t \leq \eta_s \leq 0$}
For this region we have to make a different ansatz, valid for $\eta_t, \eta_s \leq 0$, with $\Gamma_{\text{walk}}(\eta_s,\eta_t,g) \propto (1+\eta_s) (1+\eta_t)$.
This correctly vanishes at $\eta_s = \eta_t = -1$ and, matching it with \eqref{eq:all_loop_sudakov_different_st_intermediate} at $\eta_s = \eta_t = 0$, the proportionality coefficient is fixed to be $\Gamma_{\text{cusp}}/2$. 

\subsubsection*{The all-loop prediction}
The all-loop prediction for $\Gamma_{\text{walk}}(\eta_s,\eta_t,g)$ in the $\eta_t < \eta_s$ region is
\begin{equation}
\begin{cases}
0, & \eta_t \leq \eta_s \leq 0, \,\, \eta_t \leq -1,\\
\frac{\Gamma_{\text{cusp}}(g)}{2}(1+\eta_s)(1+\eta_t), & -1 \leq \eta_t \leq \eta_s \leq 0,\\
\frac{\Gamma_{\text{cusp}}(g)}{2}+\gamma(g)(\eta_s+\eta_t) + \left(\Gamma_{\text{oct}}(g) - \frac{\Gamma_{\text{cusp}}(g)}{2}-2\gamma(g)\right) \eta_s\eta_t,
& 0 \leq \eta_t \leq \eta_s \leq 1, \\[4pt]
\left(\frac{\Gamma_{\text{cusp}}(g)}{2} + \gamma(g) \right) \eta_s (1-\eta_t) + \frac{\Gamma_{\text{oct}}(g)}{4} \frac{\eta_t}{\eta_s} (1+\eta_s)^2 + \Tilde{\gamma}(g)(\eta_s-1)(\eta_t-1),
& 0 \leq \eta_t \leq 1<\eta_s, \\[8pt]
\frac{\Gamma_{\text{oct}}(g)}{2}\left(1+\dfrac{\eta_t}{2\eta_s}+\dfrac{\eta_s}{2\eta_t}\right),
& 1 \leq \eta_t \leq \eta_s .
\end{cases}
\label{eq:Gamma-walk-etas-finiteg}
\end{equation}
which reduces to \eqref{eq:Gamma-walk-etas-ordered} when using
\begin{equation}
\Gamma_{\text{cusp}}^{(1)} = 4 g^2\,, \quad \Gamma_{\text{oct}}^{(1)} = 4g^2\,, \quad \gamma^{(1)} = 2 g^2\,, \quad \Tilde{\gamma}^{(1)} = 2 g^2.
\label{eq:oneloopsafunc}
\end{equation}

\section{Conclusions}

In this paper, we studied the interpolation of the Sudakov form factor and the four-point scattering amplitude on the Coulomb branch of $\mathcal{N} = 4$ SYM  between the on-shell regime (internal mass $m \neq 0$, external mass $M=0$) controlled by the cusp anomalous dimension, and the off-shell regime (internal mass $m = 0$, external mass $M \neq 0$) controlled by the octagon. To interpolate between the two we introduced a walking parameter, see Figure \ref{fig:walking_summary_ff} and Figure \ref{fig:walking_summary}. To study observables with Euclidean virtuality $p_i^2=-M^2<0$, we effectively analytically continue the angles on $S^5$ to imaginary values, see e.g. \cite{Correa:2012nk,Correa:2012hh,Gromov:2016rrp}. The corresponding walking anomalous dimension interpolates between the cusp anomalous dimension and the octagon. Our prediction for the form factor is given in \eqref{eq:ff_all_loop}, and for the four-point scattering amplitude in \eqref{eq:Gamma-walk-etas-finiteg}.

The results depend on four functions of the 't Hooft coupling: the familiar 
$\Gamma_{\rm cusp}$, $\Gamma_{\rm oct}$, and two new functions $\gamma, 
\tilde\gamma$, which are not known at present. For the form factor, only 
$\gamma(g)$ appears, and we calculated it up to two loops, see 
\eqref{eq:gammaff2l}. The four-point amplitude depends on an additional 
unknown function $\tilde\gamma(g)$ that we calculated up to one loop, see 
\eqref{eq:oneloopsafunc}.

It is well-known that the on- and off-shell regimes receive contributions from different pinch surfaces according to Libby-Sterman infrared power counting \cite{Libby:1978qf}. In other words, they are governed by different effective theories. The on-shell case receives leading contributions from the collinear and soft modes $k_{\rm s} \sim m$, while the off-shell case is driven instead by the collinear and ultrasoft ones $k_{\rm us} \sim M^2/Q$. The presence of the ultrasoft region threatens the potential applicability of perturbative techniques in QCD due to the coupling constant growth in the infrared. What we uncovered, however, is that the soft-collinear effective theory is valid over a very wide region of mass/virtuality parameter space as determined by the $y$-parameter. It determines the walking behavior of the form factor and four-leg amplitude. The ultrasoft-collinear theory, on the other hand, sets in only on the `shoulder' of the asymptotic limit $y \ll 1/\omega$. This gives hope for future robust QCD applications relying on the factorization of soft-collinear modes from the hard remainder, but it shows an onset of nonperturbative physics as one increases off-shellness of partonic processes.

It is interesting to point out that the nature of collinear modes in the above two effective theories is quite different. Namely, in the soft-collinear case, even with non-vanishing masses and virtualities, factorization of full infrared observables into independent collinear functions is impossible without the use of supplemental regulators in addition to the conventional dimensional one. This is the consequence of the presence of well-known rapidity divergences. They are an artifact of the leading power expansion in the corresponding pinch regions. When these are mapped into the operator language, they are intimately related to the infinite length of light-like Wilson lines defining corresponding jet factors. Once proper regulators are introduced in the factorization formalism, they can be made well-defined \cite{Collins:2011zzd}. On the other hand, for the ultrasoft-collinear theory, these do not arise at all since all Wilson lines are in fact cut off by the inverse off-shellness, making them finite \cite{Belitsky:2024yag}. Another distinguished feature of the ultrasoft-collinear theory is the lack of factorization of ultrasoft and collinear modes from each other. This is, however, hardly a limitation since there is no physical process where they can be measured separately. Thus, one can rely on the protected hard factorization to evaluate physical observables with both ultrasoft and collinear modes included in separate infrared factors. 

Our results suggest several natural directions for further study.
\begin{itemize}
\item Extending the form-factor analysis to three loops should be within reach
using the same techniques \cite{Belitsky:2025bez}. Similarly, the two-loop
analysis of the scattering amplitude should be accessible. 
\item It would be very interesting to perform a strong-coupling analysis along
the lines of \cite{Alday:2007hr,Alday:2007he}. Such an analysis
faces an immediate puzzle: the strong-coupling limit of the octagon
anomalous dimension differs from that of the cusp anomalous dimension by
an additional factor of $\pi^{-1}$,
\begin{align}
\Gamma_{\rm cusp}\Big|_{g \to \infty} = 2g
\, , \qquad
\Gamma_{\rm oct}\Big|_{g \to \infty} = \frac{4g}{\pi}
\, .
\end{align}
How does this factor arise from a classical string solution?
\item Both $\Gamma_{\rm cusp}(g)$ and $\Gamma_{\rm oct}(g)$ can be computed at finite 't Hooft coupling using integrability of $\mathcal{N} = 4$ SYM. The cusp can be extracted by the so-called BES kernel \cite{Beisert:2006ez}, while the octagon, first computed at finite coupling in \cite{Belitsky:2019fan}, can be computed from a deformation of the BES kernel. In particular, this deformation of the BES kernel, controlled by a parameter $\alpha$, was defined in \cite{Basso:2020xts}. Different values of $\alpha$ capture different anomalous dimensions, for example the octagon for $\alpha = 0$ and the cusp for $\alpha = \pi/4$. This allows to define a tilted anomalous dimension $\Gamma_{\alpha}(g)$, which interpolates between different anomalous dimensions. It is natural to then try to connect the walking anomalous dimension with $\Gamma_{\alpha}(g)$, but the two are different and the connection is not obvious. In particular, the weak-coupling expansion of $\Gamma_{\alpha}(g)$ contains higher powers of $\cos \alpha, \sin \alpha$ as the loop order increases, while the walking anomalous dimension contains at most $\eta^2$ terms. We checked that, matching one and two-loops and then using the conjectured relation between $\eta$ and $\alpha$, the tilted cusp result does not reproduce, at three and higher loops, our prediction for the walking.

It is still an open question whether a connection of some sort between the two can be found and if the walking anomalous dimension can be computed using integrability of $\mathcal{N}= 4$ SYM.

\item In this work, we studied the observables in the Euclidean region,
including the spacelike external virtualities
$p_i^2=-M^2<0$. From the perspective of ${\cal N}=4$ SYM, this
involves analytically continuing the angles on $S^5$ to imaginary
values. This analytic continuation has been considered previously,
see, for example, \cite{Correa:2012nk,Correa:2012hh,Gromov:2016rrp}.
A natural direction for future work is to extend our analysis to
timelike virtualities, $p_i^2>0$, where anomalous thresholds are expected
to play an important role \cite{Correia:2022dcu}. It would also be very
interesting to generalize the finite-coupling bootstrap analysis  of
Coulomb-branch scattering in ${\cal N}=4$ SYM \cite{Alday:2025pmg} to
nonzero external masses and virtualities.

\item High-energy hadron-hadron collisions with production of heavy vector bosons and/or Higgs at small transverse momentum are sensitive to the off-shellness of the initial-state partons that induce the hard process \cite{Catani:1990xk,Collins:1991ty,Watt:2003vf}. It would be interesting to construct factorization theorems for these cross sections and study their `walking' to the collinear regime \cite{Collins:1984kg}, where colliding partons can be regarded as massless.
\end{itemize}

\section*{Acknowledgments}

We would like to thank Frank Coronado for useful discussions. This project
has received funding from the European Research Council (ERC) under the European
Union’s Horizon 2020 research and innovation programme (grant agreement number 949077). The work
of L.F.A. is partially supported by the STFC grant ST/T000864/1. KH is supported by the European Research Council (ERC) under the European Union’s Horizon 2020 research and innovation programme (grant agreement number 101115511).

\newpage
\appendix

\section{Detailed analysis of the regions of the one-loop amplitude for $s = t$}
\label{sec:detailed_analysis_amplitude}

In this appendix we present a detailed analysis of the different regions contributing to the double-logarithmic coefficient of the one-loop amplitude $\mathcal{M}^{(1)}(s,t)$ in the double-scaling limit \eqref{eq:double_scaling_limit} for $s = t$. We consider separately the two cases $y < 1$ and $y > 1$. 

\subsection{Case $y < 1$}

In this case, equivalent to $\eta \geq  0$, the dispersion representation of the one-loop amplitude is
\begin{equation}
    \mathcal{M}^{(1)}(s,t) = -2\, \int_0^{\infty} dx\, \frac{\log\left(\frac{1+2 x \omega + 2 \sqrt{x \omega(1+x \omega)}}{1+4(x+1)y}\right)}{(1 + 4 x/\omega)\sqrt{x+y}\sqrt{(x+1/\omega)}}\,.
\end{equation}
The relevant scales for the integration variable $x$ that determine the different behaviors of the integrand are
\begin{equation}
    x \sim \omega\,, \quad x \sim \omega^{-1}\,, \quad x \sim y^{-1}\,, \quad x \sim y\,.
\end{equation}
Let us analyze in detail what happens for $y > \omega^{-1}$, corresponding to the octagon-to-cusp transition and for $y < \omega^{-1}$, corresponding to the octagon shoulder.

\subsubsection{Case $y > \omega^{-1}$: octagon-to-cusp transition}

For $y > \omega^{-1}$ we have the following ordering of the relevant scales
\begin{equation}
    \omega^{-1} < y < y^{-1} < \omega\,,
\end{equation}
and the integral can then be divided into the corresponding different regions. In each case we consider the proper simplification of the integrand, valid for that specific region, and identify the ones that give double logarithms.
Let us enumerate the different regions:
\begin{itemize}
\item For $0<x<\omega^{-1}$: 
\begin{equation}
\mathcal{M}^{(1)}(s,t)\Big{|}_{x < \omega^{-1}} = -2\, \int_0^{\omega^{-1}} dx\, \frac{\log\left(\frac{1+2 x \omega + 2 \sqrt{x \omega(1+x \omega)}}{1 + 4 y}\right)}{\sqrt{y}\sqrt{(x+1/\omega)}} = \mathcal{O}\left(\frac{1}{\sqrt{y \omega}}\right)\,,
\end{equation}
so this region does not contribute to the walking anomalous dimension.
\item For $\omega^{-1} < x < y$:
\begin{equation}
\mathcal{M}^{(1)}(s,t)\Big{|}_{\omega^{-1} < x < y} = -2\, \int_{\omega^{-1}}^y dx\, \frac{\log\left(\frac{4 x \omega}{1 + 4 y}\right)}{\sqrt{x+y}\sqrt{x}} = \mathcal{O}\left(\log(\omega y)\right)\,,
\end{equation}
and also this region is not relevant since it just gives single-logarithms.
\item The region $y < x < y^{-1}$ gives
\begin{equation}
\mathcal{M}^{(1)}(s,t)\Big{|}_{y < x < y^{-1}} = -2\, \int_{y}^{y^{-1}} dx\, \frac{\log\left(\frac{4 x \omega}{1 + 4 y}\right)}{x} = 4 \log y \log \omega + \mathcal{O}\left(\log(y)\right)\,,
\end{equation}
and contributes to the expression of the walking anomalous dimension.
\item Also $y^{-1} < x < \omega$
\begin{equation}
\mathcal{M}^{(1)}(s,t)\Big{|}_{y^{-1} < x < \omega} = -2\, \int_{y^{-1}}^{\omega} dx\, \frac{\log\left(\frac{4 x \omega}{4 x y}\right)}{x} = -2 \log^2\omega + 2 \log^2y\,,
\end{equation}
contains double logarithms.
\item Finally $x > \omega$:
\begin{equation}
\mathcal{M}^{(1)}(s,t)\Big{|}_{x > \omega} = -2\, \int_{\omega}^{\infty} dx\, \frac{\log\left(\frac{4 x \omega}{4 x y}\right)}{4x^2/\omega} = \mathcal{O}(\log(\omega/y))\,.
\end{equation}
\end{itemize}
The expression of the walking anomalous dimension can be extracted from the sum of double logarithms, as presented in \eqref{eq:gamma_2} in the main text.

\subsubsection{Case $y < \omega^{-1}$: the octagon shoulder}

On the octagon shoulder $y < \omega^{-1}$ the relevant scale are ordered as follows
\begin{equation}
    y < \omega^{-1} < \omega < y^{-1}\,.
\end{equation}
As done before, we separate the integral into the different regions and approximate the integrand consistently.
\begin{itemize}
\item The first region $0<x<y$
\begin{equation}
\mathcal{M}^{(1)}(s,t)\Big{|}_{x < y} = -2\, \int_0^{y} dx\, \frac{\log\left(\frac{1+2 \sqrt{x \omega}}{1+4 y}\right)}{\sqrt{y/\omega}} = \mathcal{O}\left(y \omega\right)\,,
\end{equation}
does not contribute to the expression on the shoulder.
\item For $y < x < \omega^{-1}$:
\begin{equation}
\mathcal{M}^{(1)}(s,t)\Big{|}_{y < x < \omega^{-1}} = -2\, \int_y^{\omega^{-1}} dx\,  \frac{\log\left(\frac{1+2 \sqrt{x \omega}}{1+4 y}\right)}{\sqrt{x/\omega}} = \mathcal{O}\left(1\right)\,.
\end{equation}
\item The region $\omega^{-1} < x < \omega$ is relevant, since it gives double logarithms
\begin{equation}
\mathcal{M}^{(1)}(s,t)\Big{|}_{\omega^{-1} < x < \omega} = -2\, \int_{\omega^{-1}}^{\omega} dx\, \frac{\log\left(\frac{4 x \omega}{1+4 y}\right)}{x} = -4 \log^2 \omega + \mathcal{O}\left(y\right)\,.
\end{equation}
\item The result for $\omega < x < y^{-1}$
\begin{equation}
\mathcal{M}^{(1)}(s,t)\Big{|}_{\omega < x < y^{-1}} = -2\, \int_{\omega}^{y^{-1}} dx\, \frac{\log\left(\frac{4 x \omega}{1+4 y}\right)}{4x^2/\omega} = \mathcal{O}(\log\omega)\,,
\end{equation}
does not contribute.
\item Finally, for $x > y^{-1}$:
\begin{equation}
\mathcal{M}^{(1)}(s,t)\Big{|}_{x > y^{-1}} = -2\, \int_{y^{-1}}^{\infty} dx\, \frac{\log\left(\frac{4 x \omega}{4 x y}\right)}{4x^2/\omega} = \mathcal{O}(y \omega \log(\omega/y))\,.
\end{equation}
\end{itemize}

\subsection{Case $y  > 1$}
To extract the interpolating anomalous dimension for $y > 1$, corresponding to $\eta \leq 0$, the dispersion representation of the one-loop amplitude has to be simplified with $y \gg 1$. It reads
\begin{equation}
\mathcal{M}^{(1)}\Big{|}_{y \gg 1} = -2 \int_0^{\infty} dx \, \frac{\log \left( \frac{(\sqrt{x \omega} + \sqrt{x \omega + 4 y (x+y)})^2}{4 y (x+y)} \right)}{(1 + 4 (x+y)/\omega) \sqrt{x+y} \sqrt{(x + 4y (x+y)/\omega)}}\,,
\end{equation}
with the relevant scales for the integration variable $x$ given by
\begin{equation}
    x \sim \omega\,, \quad x \sim \omega^{-1}\,, \quad x \sim y^{-1}\,, \quad x \sim y\,,
\end{equation}
as before. We proceed as in the previous section by identifying the double logarithms in the case $1 < y < \omega$, which describes the transition from the cusp to zero. The case of $y > \omega$ is not analyzed since the result for the walking anomalous dimension is just $\Gamma_{\rm walk}^{(1)}(\eta \leq -1) = 0$. 

\subsubsection{Case $1< y < \omega$:} 

The ordering of the relevant scales for the cusp-to-zero transition is
\begin{equation}
    \omega^{-1} < y^{-1} < y < \omega\,.
\end{equation}
Let us analyze in detail the different contributions.
\begin{itemize}
\item The first region $0 < x < \omega^{-1}$ gives
\begin{equation}
\mathcal{M}^{(1)}\Big{|}_{x < \omega^{-1}} = -2 \int_0^{\omega^{-1}} dx \, \frac{\log\left(1+\frac{\sqrt{x \omega}}{2y}\right)}{(1 + 4 y/\omega) \sqrt{y} \,(2 y/\sqrt{\omega})} = \mathcal{O}\left( \frac{1}{y^{5/2}\sqrt{\omega}}\right)\,.
\end{equation}

\item For the region $\omega^{-1} < x < y^{-1}$ 
\begin{equation}
\mathcal{M}^{(1)}\Big{|}_{\omega^{-1} < x < y^{-1}} = -2 \int_{\omega^{-1}}^{y^{-1}} dx \, \frac{\log\left(\frac{(\sqrt{x \omega} + \sqrt{x \omega + 4 y^2})^2}{4 y^2}\right)}{(1 + 4 y/\omega) \sqrt{y} \sqrt{x + 4y^2/\omega}} = \mathcal{O}\left( \frac{\log \omega}{y} \right)\,,
\end{equation}
and does not contribute. 

\item The region $y^{-1} < x < y$ gives
\begin{equation}
\mathcal{M}^{(1)}\Big{|}_{y^{-1} < x < y} = -2 \int_{y^{-1}}^y dx \, \frac{\log\left(\frac{(\sqrt{x \omega} + \sqrt{x \omega + 4 y^2})^2}{4 y^2}\right)}{(1 + 4 y/\omega) \sqrt{y} \sqrt{x + 4y^2/\omega}} = \mathcal{O}\left(\log \omega, \log y\right)\,,
\end{equation}

\item The region $y < x < \omega$ contributes with a double logarithm:
\begin{equation}
\mathcal{M}^{(1)}\Big{|}_{ y < x < \omega} = -2 \int_{y}^{\omega} dx \, \frac{\log\left(\frac{(\sqrt{x \omega} + \sqrt{x \omega + 4 x y})^2}{4 x y}\right)}{\sqrt{x (x + 4 x y/\omega)}} = -2 \log^2\left( \frac{\omega}{y}\right) + \mathcal{O}(y^{-1})\,.
\end{equation}
\item Finally, for $\omega < x < \infty$:
\begin{equation}
\mathcal{M}^{(1)}\Big{|}_{ \omega < x < \infty} = -2 \int_{\omega}^{\infty} dx \, \frac{\log\left(\frac{\omega}{y}\right)}{4 x^2/\omega} = \mathcal{O}\left( \log \frac{\omega}{y} \right)\,.
\end{equation}
\end{itemize}
The only contribution to the interpolating anomalous dimension comes from the region $y < x < \omega$, as also reported in the main text \eqref{eq:neg_eta_one_loop}.

\newpage
\bibliographystyle{JHEP} 

\bibliography{literature.bib}

\end{document}